\documentclass[twocolumn,superscriptaddress]{revtex4-2}

\usepackage[]{graphicx}
\usepackage{dcolumn}
\usepackage{bm}

\usepackage{amsmath,amssymb,amsfonts}
\usepackage{fancyhdr}
\usepackage{lipsum} 
\usepackage{subcaption}
\usepackage{multirow}

 \usepackage[colorlinks]{hyperref}
 \hypersetup{
     colorlinks = false,
   }

\newcommand{\comment}[1]{}

\usepackage{color}
\usepackage{babel}
\newcommand{\bea}{\begin{eqnarray}}
\newcommand{\eea}{\end{eqnarray}}

\newcommand{\be}{\begin{equation}}
\newcommand{\ee}{\end{equation}}

\begin{document}

\title[]{Multikink scattering in the $\phi^6$ model revisited}

\author{C. Adam}
\email[]{adam@fpaxp1.usc.es}
\affiliation{Departamento de F\'isica de Part\'iculas, Universidad de
Santiago de Compostela and \\
Instituto Galego de F\'isica de Altas Enerxias (IGFAE), E-15782
Santiago de Compostela, Spain}
\author{P. Dorey}
\email[]{p.e.dorey@durham.ac.uk}
\affiliation{Department of Mathematical Sciences, Durham University, UK}

\author{A. García Martín-Caro}
\email[]{alberto.martincaro@usc.es}
\affiliation{Departamento de F\'isica de Part\'iculas, Universidad de
Santiago de Compostela and \\
Instituto Galego de F\'isica de Altas Enerxias (IGFAE), E-15782
Santiago de Compostela, Spain}

\author{M. Huidobro}
\email[]{miguel.huidobro.garcia@usc.es}
\affiliation{Departamento de F\'isica de Part\'iculas, Universidad de
Santiago de Compostela and \\
Instituto Galego de F\'isica de Altas Enerxias (IGFAE), E-15782
Santiago de Compostela, Spain}

\author{K. Oles}
\email[]{katarzyna.slawinska@uj.edu.pl}
\affiliation{Institute of Theoretical Physics, Jagiellonian University,
Lojasiewicza 11, Krak\'{o}w, Poland}

\author{T. Romanczukiewicz}
\email[]{tomasz.romanczukiewicz@uj.edu.pl}
\affiliation{Institute of Theoretical Physics, Jagiellonian University,
Lojasiewicza 11, Krak\'{o}w, Poland}

 \author{Y. Shnir}
\email[]{shnir@maths.tcd.ie}
\affiliation{Institute of Physics, University of Oldenburg, Oldenburg D-26111, Germany}

\author{A. Wereszczynski}
\email[]{andrzej.wereszczynski@uj.edu.pl}
\affiliation{Institute of Theoretical Physics, Jagiellonian University,
Lojasiewicza 11, Krak\'{o}w, Poland}

\begin{abstract}

Antikink-kink ($\bar{\rm K} $$ {\rm K}$) collisions in the $\phi^6$ model exhibit resonant scattering although the $\phi^6$ kinks do not support any bound states to which energy could be transferred. In {\em Phys. Rev. Lett. {\bf 107} (2011) 091602} it was conjectured that, instead, energy is transferred to a collective bound mode of the full  $\bar{\rm K} $$ {\rm K}$ configuration. Here we present further strong evidence for this conjecture.

Further, we construct a collective coordinate model (CCM) for $\bar{\rm K} $$ {\rm K}$ scattering based on this collective bound mode trapped between the $\bar{\rm K} $$ {\rm K}$ pair which allows us to reproduce the full dynamics of $\bar{\rm K} $$ {\rm K}$ scattering with striking accuracy. We also study kink-antikink ($ {\rm K}$$\bar{\rm K} $) scattering and its description by a CCM. 
In this case a significant role of radiation is discovered.

%We present a detailed evidence that degrees of freedom (bound modes) effectively emerging in multikink states and located {\it between} solitons rather than those confined to single solitons explain the fractal structure observed in antikink-kink collision in $\phi^6$ theory. 

%This allows us to propose a robust way for construction of a restricted set of configurations, and therefore, collective coordinate model (CCM), which may capture a solitonic proces in a wide class of field theories in (1+1) dimensions in terms of finite number of moduli. This is applied to the antikink-kink scattering in $\phi^6$ model where we reproduce the full theory dynamics with a striking accuracy. The kink-antikink is also studied. In this case a significant role of radiation is discovered.  
\end{abstract}
\maketitle

%%%%%%%%%%%%%%%%%%%%%%%%%%%%%
\section{Introduction}
%%%%%%%%%%%%%%%%%%%%%%%%%%%%%
Topological solitons are spatially localized, stable solutions of nonlinear field equations which carry a nonzero topological charge \cite{R, SM, Shnir}.
The quantitative and sometimes even the qualitative understanding of their interactions
is far from satisfactory in many cases.
 Since solitons exist both in fundamental theories (e.g., sphalerons or monopoles in the electroweak theory) as well as in numerous effective models, a comprehensive understanding of their interactions  is vital not only for a deeper insight into the mathematical structure of the theories but also for applications.  

There are three main contributions to solitonic interactions. 

Firstly, two solitons at a finite distance can act on each other with a {\it static force}. This force can be attractive (as typically happens for a kink-antikink pair) or repulsive. In some particular cases, the so-called  Bogomonly-Prasad-Sommerfeld (BPS) models \cite{Bo}, there is no static force between the constituent solitons in a multi-soliton state \cite{JT}, \cite{Ma8}. There are famous examples of this in higher dimensions  such as the Abelian Higgs model at critical coupling or BPS monopoles \cite{AH, Sam, nick}, but they also exist in impurity-deformed \cite{BPS_imp-1, solv} or multi-field models in (1+1) dimensions \cite{Bazeia, Izq-1, Izq-2, 2field-SW}. 

Secondly, the dynamics may be significantly affected by the excitation of {\it internal} degrees of freedom (DoF). These are often massive normal or quasinormal modes supported by solitons, found in  linear perturbation theory. Later on we will see, however, that other possibilities also play  a very important role. A well-known example for the impact of internal DoF on soliton dynamics is the so-called {\it resonance phenomenon} which is responsible for the fractal structure observed in the final state production in kink-antikink 
($\bar{\rm K}$ ${\rm K}$) 
collisions in various (1+1) dimensional solitonic models \cite{Sug,CSW}, \cite{good-1}. Here, during the collision, the initial kinetic energy of the incoming kink and antikink is transferred to internal DoF. Then, it can be transferred back to the kinetic DoF, allowing for the re-appearance of the solitons in the final state. However, it is also possible that a significant fraction can be kept in the internal DoF, not allowing the solitons to escape, which eventually leads to their complete annihilation. Both scenarios occur in a chaotic manner, resulting in the well-known fractal pattern. This mechanism has been recently confirmed in  ${\rm K}$ $\bar{\rm K}$ scattering in the $\phi^4$ theory by a derivation of a {\it collective coordinate model} (CCM) based on the zero mode (kinetic DoF) and a massive normal mode, the so-called shape mode \cite{MORW}. In fact, the resulting CCM reproduces the fractal structure with quite good agreement. This has been further improved by considering, instead, a tower of Derrick modes, that is, scaling deformations. This framework, the so-called {\it perturbative relativistic CCM} not only allows to incorporate relativistic corrections in a systematic, well defined perturbative fashion, but it also provides an arbitrary number of collective coordinates, which results in a more accurate description of solitonic processes. E.g., the inclusion of the two lowest Derrick modes already led to a CCM which describes the ${\rm K}$ $\bar{\rm K}$ process in the $\phi^4$ model with a few percent precision \cite{AMORW}. 

Finally, in multi-soliton processes the constituent solitons interact with radiation which is easily produced during the scattering. Radiation is an excitation of the continuous spectrum (scattering modes) which is not confined to the solitons but propagates in the full space. The interaction between solitons and radiation is a very important but, at the same time, a very complicated topic where the famous {\it soliton resolution conjecture} still waits to be proven \cite{resol1,resol2}. Due to nonlinearities, radiation may have a rather surprising effect on soliton motion like, e.g., the {\it negative radiation pressure}, e.g., \cite{neg-rad}. Note that, in contrast to internal DoF, radiation provides a channel in which the energy can escape from the solitons. 

All these three types of interactions are obviously coupled to each other, rendering the rigorous analysis extremely difficult even in the simplest case of ${\rm K}$ $\bar{\rm K} $ collisions in (1+1) dimensions. Note that the use of Derrick modes may probably take into account both the massive normal modes as well as the continuous spectrum, at least partially. This is because higher Derrick modes have frequencies above the mass threshold \cite{AMORW}. On the other hand, Derrick modes, of course,  do not represent proper radiation, since they are confined to solitons, although higher Derrick modes are more and more widely spread out. 

Although the recent progress in the explanation of the ${\rm K}$ $\bar{\rm K}$ collisions in the $\phi^4$ model \cite{MORW, AMORW} is a significant step forward in the understanding of the dynamics of solitons, it is just a beginning. There are many kink models whose dynamics are still understood to a rather unsatisfactory extent, see, e.g., deformations of the $\phi^6$ model \cite{gani-deformed-phi6, S1, T1}, higher order potentials \cite{higher-nick, kev, Decker-1, Mohammadi-2}, double sine-Gordon \cite{2sg, gani-2sg, simas-2, Mohammadi-1} and other models \cite{other-simas-1, other-Izq-1, other-Izq-2, bazeia-1, bazeia-2, simas-1, simas-3}.  

In the present paper we revisit the antikink-kink ($\bar{\rm K} $$ {\rm K}$) and kink-antikink ($ {\rm K}$$\bar{\rm K} $) collisions in the $\phi^6$ model \cite{Tr, W1, gani-phi6, W2}. This is the simplest theory which goes beyond $\phi^4$ model and which, for the $\bar{\rm K}$ ${\rm K}$ case, exhibits a fractal structure in the final state formation despite the nonexistence of shape modes hosted by a single (anti)kink. However, as originally proposed in \cite{Tr}, the resonant phenomenon can be triggered by dynamically created modes trapped between the colliding solitons. The fact that solitonic dynamics can be significantly affected by temporal variations of the mode structure has been recently appreciated, see, for instance, the spectral wall phenomenon \cite{spectral-wall} or the role played by unstable solutions, i.e., {\it sphalerons}, occurring during a multi-kink evolution \cite{sphaleron}. 

On the other hand, there is no corresponding fractal structure in the ${\rm K}$ $\bar{\rm K}$ case as no trapped modes exist. 

Here, we provide a detailed exploration of $\bar{\rm K} $$ {\rm K}$ and 
$ {\rm K}$$\bar{\rm K} $ collisions based on full numerical simulations. Further, we shall propose strategies for the construction of CCMs for kink collisions. 
 In particular, we will present a CCM which reproduces the full numerical $\bar{\rm K}$ ${\rm K}$ collisions with a high precision. We will also discover some interesting difficulties concerning the construction of a CCM for the ${\rm K}$ $\bar{\rm K} $ collisions. Here the inclusion of radiation effects seems to be inevitable.

%%%%%%%%%%%%%%%%%%%%%%%%%%%%%
\section{Road map to restricted set of configurations}
%%%%%%%%%%%%%%%%%%%%%%%%%%%%%
Let us consider a scalar field theory in (1+1) dimensions 
\be
L[\phi]=\int_{-\infty}^\infty dx \left( \frac{1}{2} \phi_t^2 - \frac{1}{2} \phi_x^2
  - U(\phi) \right)\, ,
\ee
where the field theoretical potential $U$ has at least two vacua, $\phi_v^{(1)}>\phi_v^{(2)}$. This guarantees the existence of a topological (anti)kink $\Phi_{K(\bar{K})}$ which interpolates between the vacua. In principle, we do not have to assume that the vacua are approached quadratically. This means that the mass of small perturbations around a vacuum, i.e., the {\it meson mass}, can take any value $0\leq m_{1,2} \leq \infty$. A finite, non-zero mass means a quadratic approach, as for example in the $\phi^4$ and $\phi^6$ models 
\be
U_{\phi^4}= \frac{1}{2} (1-\phi^2)^2, \;\;\; U_{\phi^6}= \frac{1}{2} \phi^2(1-\phi^2)^2.
\ee
On the other hand, the $\phi^8$ potential
\be
U_{\phi^8}= \frac{1}{2} \phi^4(1-\phi^2)^2
\ee
has a vacuum at $\phi_v=0$ which is approached with a higher than quadratic power. Therefore, the mass of mesons vanishes and there is no mass gap in this vacuum. On the other hand a formally infinite mass of vacuum excitations is characteristic for compactons, which are solitons with finite support \cite{arodz}.

The {\it collective coordinate model} (CCM) approach is a standard approach which allows for a semi-analytical treatment of the dynamics of topological solitons and, therefore, offers a tool for the explanation of a variety of phenomena occurring in multi-soliton collisions, see \cite{SM} for a review. In this framework, the infinitely many DoF of the original field $\phi$ whose dynamics is governed by the Lagrangian $L[\phi]$  are replaced by a finite number of DoF arising from the truncation of the space of fields to a {\it restricted} set of static configurations 
\be
\mathcal{M}=\{ \Phi(x; \, X^i), \, i=1..N \} \, .
\ee
$\mathcal{M}$ is called the moduli space and it should contain configurations which capture the main features of a given solitonic process. Its dimension, $N$, depends of the number of parameters, i.e., {\it moduli}, $X^i$, which are then promoted to time dependent variables $X^i(t)$. The configurations $\Phi$ are inserted into the original Lagrangian ${L}[\phi]$ and after performing spatial integration we obtain an effective CCM theory
\be
L({\bf X}) = \int_{-\infty}^\infty \mathcal{L} [\Phi(x; {\bf X})] dx = \frac{1}{2}
g_{ij}({\bf X}) \dot{X}^i \dot{X}^j - V({\bf X}) \,,
\label{eff-lag}
\ee
where
\be
g_{ij}({\bf X})=\int_{-\infty}^\infty \frac{\partial \Phi}
{\partial X^i} \frac{\partial \Phi}{\partial X^j} \, dx \label{metric}
\ee
is the metric on $\mathcal{M}$ while
\be
V({\bf X})=\int_{-\infty}^\infty \left( \frac{1}{2}
\left( \frac{\partial \Phi}{\partial x}
\right)^2 + U(\Phi) \right) \, dx \label{potential}
\ee
is the positive-definite potential.

The essential ingredient is, therefore, the correct choice of the restricted set of configurations. Unfortunately, except for the BPS models, there is no canonical way to construct this set.  An obvious criterion is the following:
\begin{itemize}
\item[{\it (i)}] {\it The restricted configurations should reproduce, as much as possible, the actual field profiles occurring during a given process.} 
\end{itemize}

We begin with the simplest single kink sector. As a kink is a solution of a Poincar\'e invariant theory, it enjoys translational symmetry and, therefore, can be located at any spatial point $a$. This leads to a whole family of energetically equivalent kinks $\Phi_K(x;a)=\Phi_K(x-a)$. This is also related to the fact that a kink is a static solution of a first order equation, the so-called {\it Bogomolny equation},
\be
\phi_x = \pm \sqrt{2U},
\ee 
which implies the appearance of one integration constant, the modulus $a$. Of course, the transition between energetically equivalent solutions costs an arbitrarily small amount of energy. Therefore, the modulus $a$ can be identified with the presence of a zero mode and its change describes a kinetic DoF, i.e., the rigid motion of the kink. 

In the next step, one may also include massive excitations. They can be vibrational modes obtained in the linear perturbation analysis (normal modes or quasinormal modes). Here, we perturb the kink solution by a small deformation, $\phi(x,t) = \Phi_K(x;a) + \eta^K (x,t;a)$. If inserted into the equation of motion, this leads to the linear Schroedinger-type equation
\be
\left. \left( \frac{d^2}{dx^2}  - \frac{d^2U}{d\phi^2}\right|_{\Phi_K(x;a)} \right) \eta^K(x;a) = - \omega^2\eta^K (x;a),
\ee
where we assumed periodic time dependence of the perturbation, $\eta^K(x,t;a)=\eta^K(x;a) e^{i\omega t}.$  Normal modes require $\omega \in \mathbb{R}$ while for quasi-normal modes $\omega$ has a nontrivial imaginary part. 

However, this is not the only possible choice. For example, it has recently been proposed to use Derrick modes arising in a scaling perturbation \cite{AMORW}. This not only introduces an arbitrary number of modes (hence, collective coordinates) but also recovers, in a perturbative fashion, the Lorentz contraction of solitons. In addition, higher Derrick modes have frequencies above the mass threshold (the mass of the meson squared), which may take into account some features of radiation. 

Both choices simply represent internal DoF of a kink. As a result, we arrive at the following restricted set of configurations 
\be
\mathcal{M}_{K}=\{ \Phi_{K}(x; a)+ \sum_{i=1}^NX^i  \eta^K_i(x,a) \} \, ,
\ee
where for concreteness we choose the normal modes $\eta^K_i(x;a)$ hosted by a kink. $X^i$ are the amplitudes of the modes and serve as new moduli. Note that due to the Poincare invariance $\eta^K_i(x,a)=\eta^K_i(x-a)$ and the internal DoF are confined to the kink. The construction for the antikink is identical.

Now, we can move to a multi soliton processes like antikink-kink or kink-antikink collisions. The starting point is the simplest, one-parameter restricted set of configurations, $\{ \Phi(x;a) \}$,  which should provide the crudest description of the considered process. This one-dimensional moduli space is not meant to give a good approximation but rather a correct arena in which the multi-kink scattering happens. This means the following.
\begin{itemize}
\item[{\it(ii)}]  {\it The one-parameter set $\{ \Phi(x;a) \}$ should include the initial and final states occurring in the process.}
\end{itemize}
In other words, $\lim_{a\to a_\pm } \Phi(x;a) =\Phi_{in(out)}$, where $a_\pm$ are some of the values of the modulus $a$. It is important to fulfill this condition, which is more restrictive than the condition of the correct topology. In many cases, it is enough to take a simple sum of the scattered (initial) solitons (modulo an additive constant setting the proper value of the vacuum). This happens, e.g., for $ {\rm K}$$\bar{\rm K} $ collisions in the $\phi^4$ model, where $\Phi(x;a) = \Phi_K(x+a) +\Phi_{\bar{K}}(x-a)$. This will also be a valid construction for antikink-kink collisions in the $\phi^6$ model. Nonetheless, we will present an example where such a naive sum leads to configurations which do not interpolate between the initial and final states, see the kink-antikink scattering in $\phi^6$ theory. 

Let us assume that we have found the correct one-dimensional moduli space obeying condition {\em (ii)} above. For the sake of simplicity we also assume that the simple sum of the initial states is a good choice. Now we want to include further DoF. 
The most straightforward choice for the restricted set of configurations is simply the naive sum of single soliton sets. For example, for an antikink-kink collision this results in the following moduli space
\bea
& &\mathcal{M}_{\bar{K}K}=\mathcal{M}_{\bar{K}} \cup \mathcal{M}_{K} =\label{naive-M} \\
&&\{ \Phi_{\bar{K}}(x; -a)+\Phi_{K}(x; a)+ \sum_{i=1}^NX^i  \left( \eta^{\bar{K}}_i(x,-a) +\eta^K_i(x,a) \right)\}, \nonumber 
\eea
where we restrict our consideration to a symmetric scattering. Here the antikink and kink are located at $-a$ and $+a$, respectively. In general, this choice may lead to the appearance of apparent singularities on the moduli space. Usually, they can be removed by an appropriate redefinition (or extension) of the collective coordinates \cite{MORW-moduli}. Another important observation is that, by construction, all internal DoF are {\it confined} to the constituent solitons. 

\begin{figure}
 \includegraphics[width=0.45\columnwidth]{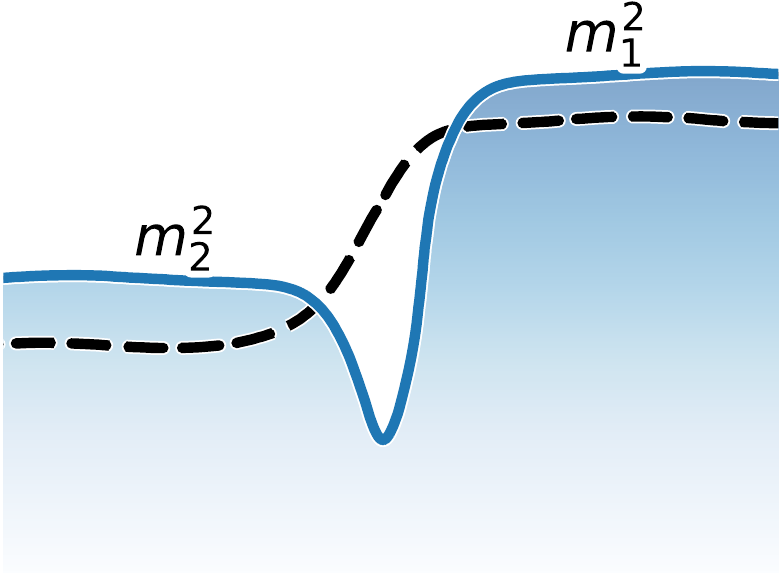}
 \includegraphics[width=0.45\columnwidth]{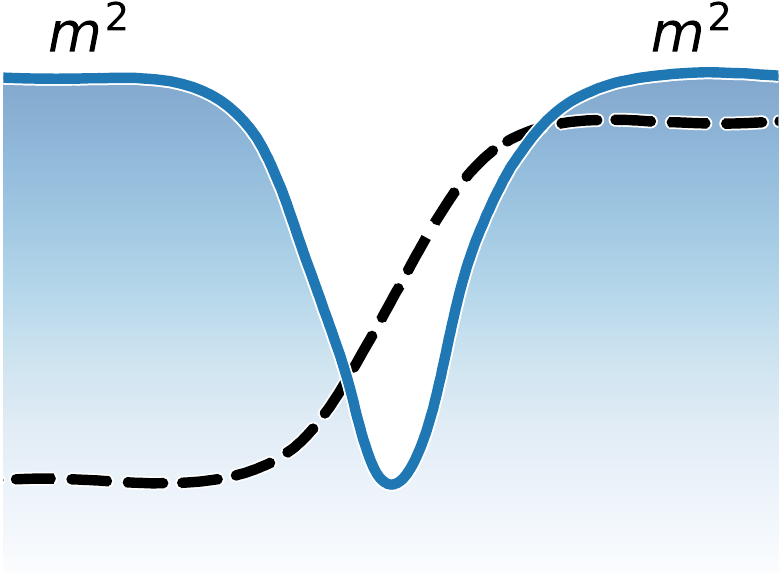}
 \caption{Schematic plot of the single kink eigen-problem potential $U_{\phi\phi}\left|\Phi_K \right.$. Left: asymmetric case, e.g., in $\phi^6$ model. Right: symmetric case, e.g., in $\phi^4$ model.} \label{kink_fig}
 \end{figure}
\begin{figure}
 \includegraphics[width=0.45\columnwidth]{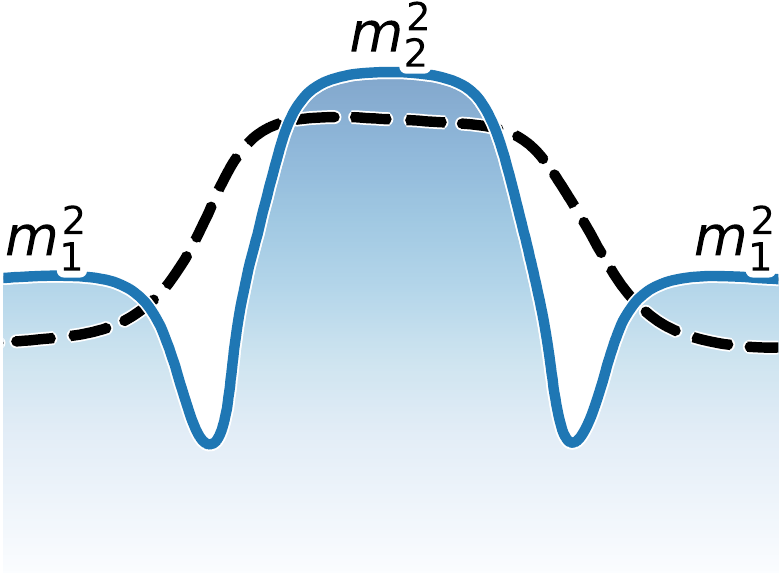}
 \includegraphics[width=0.45\columnwidth]{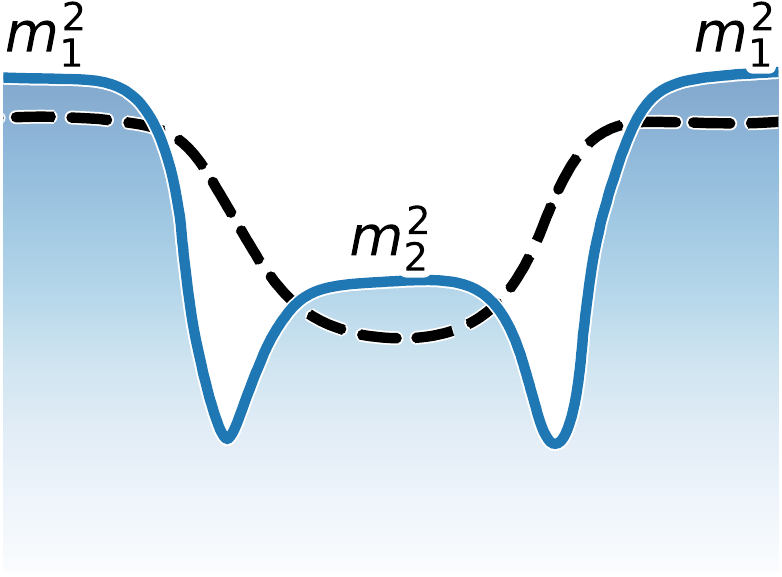}
 \caption{Schematic plot of the two-kink eigen-problem potential for the asymmetric solitons. Left: KAK case, $U_{\phi\phi}\left|\Phi_{KAK}\right.$. Right: AKK case, $U_{\phi\phi}\left|\Phi_{AKK}\right.$.}\label{kak_fig}
 \end{figure}
 
Although this framework very successfully reproduced, e.g., kink-antikink collisions in the $\phi^4$ model, especially in the case where Derrick modes are used, in general it may miss an important detail, namely the possible existence of modes which are not confined to individual solitons. 
 
To see this possibility, we consider again a kink and its linear perturbation. In general, the effective potential in the eigen-problem can tend to two different values at $x \to \pm \infty$. This corresponds to two different meson masses in the two vacua, $m_1< m_2$. This is schematically plotted in Fig. \ref{kink_fig}, left panel. This happens, for example, in $\phi^6$ theory. The mass at the vacuum $\phi=0$ is $m_2=1$ while at the vacuum $\phi=1$ it takes the bigger value $m_1=2$. For the $\phi^4$ model the masses are identical and we have a symmetric situation, see Fig. \ref{kink_fig}, right panel. 

Now,  if we consider  a two-soliton configuration built as a naive sum of single soliton states, we find two qualitatively distinct possibilities for well-separated constituent solitons. The single soliton effective potentials can be joined by a plateau with the bigger mass threshold $m_2$, see Fig. \ref{kak_fig}, left panel. This corresponds to a kink-antikink state, e.g., in the $\phi^6$ model. Obviously, the resulting modes are simply a sum of modes of each soliton. Furthermore, they are still confined to the solitons. However, if the effective potentials are connected by a plateau with smaller mass $m_1$, some new modes may show up, see Fig. \ref{kak_fig}, right panel. These modes are not confined to the colliding solitons but are delocalized in the space between them. These are new, {\it two-soliton}, {\it trapped} or {\it delocalized modes} and they play an important role in a scattering process.  
We call them delocalized modes because, in contrast to the usual normal modes, they are not bounded to individual solitons but, instead, extend to the whole space between the colliding kinks.
Exactly this situation occurs in antikink-kink collisions in the $\phi^6$ model where, in addition, single solitons do not host any normal modes. Therefore, the observed fractal structure in the final state formation was attributed to the resonant phenomenon between the kinetic DoF and the two-soliton modes \cite{Tr}. A similar mechanism was later considered in the case of the $\phi^8$ model \cite{gani-phi}. 

This leads to an improved proposal for the restricted set of configurations.
\begin{itemize}
\item[{\it (iii)}] {\it The multi-parameter restricted set of configurations, $\{ \Phi(x;a,X^1,..,X^N)\}$, should include the internal DoF (internal modes) arising in the perturbation theory of the corresponding multi-soliton configuration $\Phi(x;a)$.}
\end{itemize}
Concretely, for antikink-kink collisions this leads to the following moduli space
\bea
& &\tilde{\mathcal{M}}_{\bar{K}K}=\{ \Phi(x;a,X^1,..,X^N)\} \\
&&\{ \Phi_{\bar{K}}(x; -a)+\Phi_{K}(x; a)+ \sum_{i=1}^NX^i  \eta^{\bar{K}K}_i(x,a) \},  \nonumber
\eea
where $\eta^{\bar{K}K}_i(x,a)$ contains both the localized and the {\it delocalized} modes obtained for the effective potential 
\be
\left. \frac{d^2U}{d\phi^2}\right|_{\Phi_{\bar{K}}(x; -a)+\Phi_{K}(x; a)}.
\label{AKK-pot-eff}
\ee
Here we use our assumption that the one-dimensional moduli space is spanned by a simple sum of the antikink and kink, $\Phi(x;a)=\Phi_{\bar{K}}(x; -a)+\Phi_{K}(x; a)$. In general, also the localized soliton modes should be added as in (\ref{naive-M}). Indeed, if there are no delocalized modes, such a naive superposition works quite well. Therefore, this improved set contains the set obtained by naive superposition, $\tilde{\mathcal{M}} \supset \mathcal{M}$. 

Several comments are in order. Firstly, 
while the localized soliton modes are assumed to have the same form at any intersoliton distance $2a$, this is not the case for the delocalized modes. On the contrary, by construction their form as well as their number changes as the solitons approach each other. In particular, at a certain distance between the solitons a delocalized mode hits the mass threshold and becomes a non-normalizable mode. This means that it ceases to be a valid DoF. Such a change of the structure of the modes may potentially have a very nontrivial impact on the kink dynamics, which has to be carefully taken into account in the restricted set of configurations. This fact also is behind the spectral wall phenomenon \cite{spectral-wall, sw-1}. 

Secondly, it may be necessary to include radiation, which in particular cases can strongly affect the dynamics. Interestingly, the delocalized modes can be identified with {\it localized} radiation with a frequency below the higher mass threshold. Indeed, for sufficiently separated solitons, an initial perturbation localized at the origin will decay into radiation identically as if being created in the vacuum with lower meson mass. The fraction of radiation with $\omega < m_1$ will be trapped and eventually can excite delocalized modes. The number of these modes goes to infinity and the energy level separation decreases as the intersoliton distance grows, and effectively we flow into the continuous spectrum. This is a rather remarkable mode-radiation duality. 

In the next section we will present further arguments that delocalized, trapped modes are indeed responsible for the fractal structure in $\bar{\rm K}{\rm K}$ collisions in the $\phi^6$ model, as originally proposed in \cite{Tr}. 

Later on, we will consider ${\rm K}\bar{\rm K}$ collisions where no delocalized modes exist. As there are also no single soliton massive normal modes, the dynamics seems to be governed by radiation, which quite efficiently transfers the energy from the kinetic DoF leading to a fast annihilation. This will lead to a new challenge for a description within a CCM. 
%%%%%%%%%%%%%%%%%%%%%%%%%%%%%
\section{$\bar{\rm K}{\rm K}$ collisions in the $\phi^6$ model revisited}
%%%%%%%%%%%%%%%%%%%%%%%%%%%%%
The $\phi^6$ model is defined by the following Lagrangian 
\be
L_{\phi^6}[\phi]= \int_{-\infty}^\infty \left( \frac{1 }{2} \phi_t^2 - \frac{1}{2}\phi_x^2 - \frac{1}{2} \phi^2\left(1-\phi^2\right)^2 \right) dx
\ee
and has a static kink
\be
\Phi_{K} (x;a)\equiv \phi_{(0,1)}(x;a) =\sqrt{\frac{1+\tanh(x-a)}{2}}
\ee
which interpolates between two vacua: $\Phi_K(x=-\infty)=0$ and $\Phi_K(x=+\infty)=1$. The antikink joins the same vacua but in opposite order and reads
\be
\Phi_{\bar{K}} (x;a) \equiv \phi_{(1,0)}(x;a) =\sqrt{\frac{1-\tanh(x-a)}{2}}
\ee
In contrast to $\phi^4$ solitons, kink and antikink are not related by a simple multiplication by $(-1)$. There are also mirror kinks $\Phi^*_K$ and antikinks $\Phi^*_{\bar{K}}$ interpolating between the vacua at $-1$ and $0$. They are $\Phi^*_K (x;a) \equiv \phi_{(-1,0)}(x;a) = - \phi_{(0,1)}(-x;-a)$ and 
$\Phi^*_{\bar{K}} (x;a) \equiv \phi_{(0,-1)}(x;a) = - \phi_{(0,1)}(x;a)$.

The solitons host only a zero mode reflecting the existence of the free parameter $a$. There are no massive normal modes. In addition, as we have already  mentioned, the mass of small perturbations around the $\phi_v=0$ vacuum is smaller than for the $\phi_v=\pm 1$ vacua. Namely, $m_0=1$ while $m_{\pm1}=2$. This leads to an asymmetric effective potential in the linear perturbation problem. 
 
 \begin{figure}
 \includegraphics[width=1.00\columnwidth]{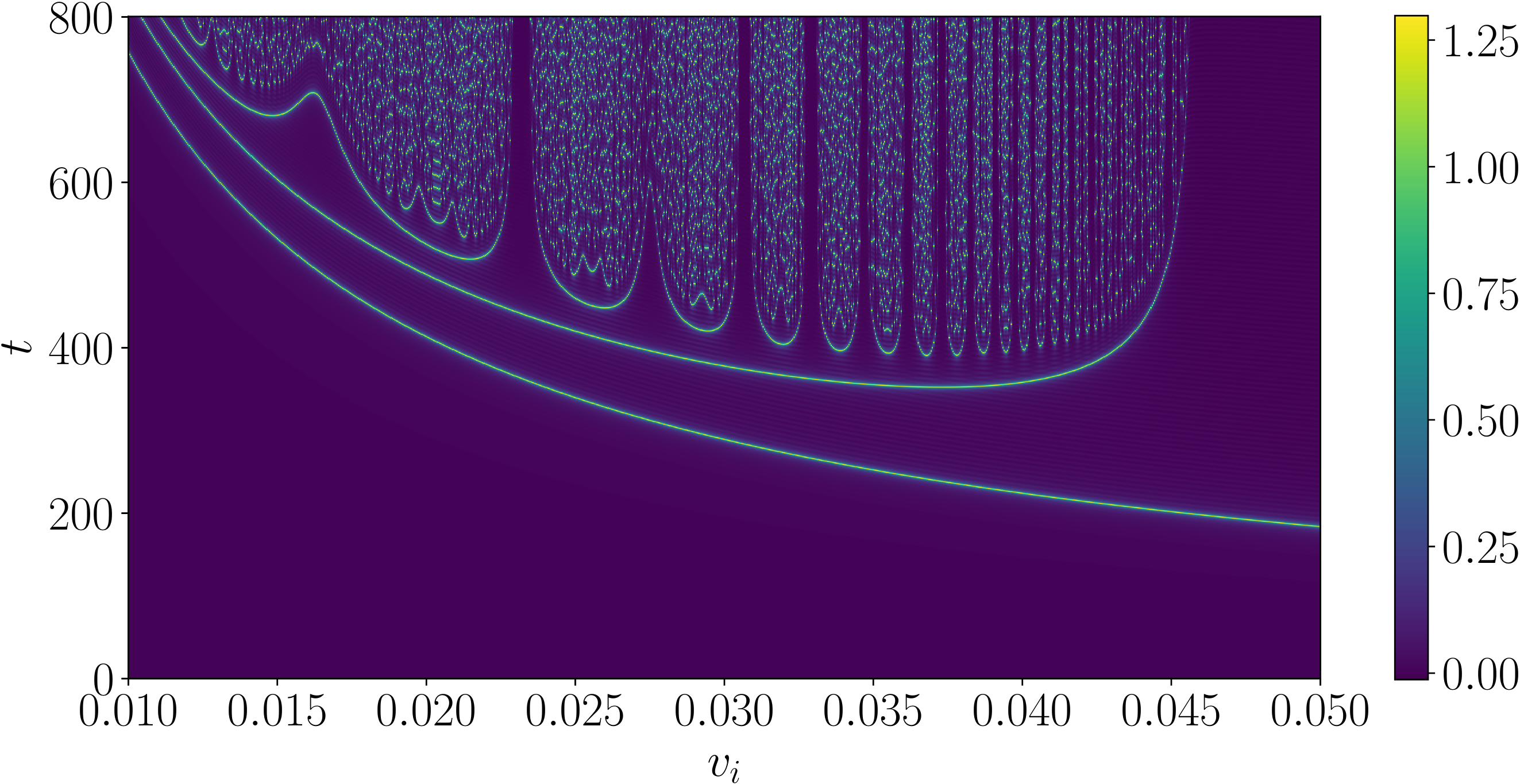}
 \caption{Time dependence of the value of the field at the origin, $\phi(x=0,t)$ for various initial velocities $v_{in}$ in the $\bar{\rm K}{\rm K}$ collision in the $\phi^6$ model.}\label{akk_scan_fig}
 \end{figure}
\begin{figure}
 \includegraphics[width=1.00\columnwidth]{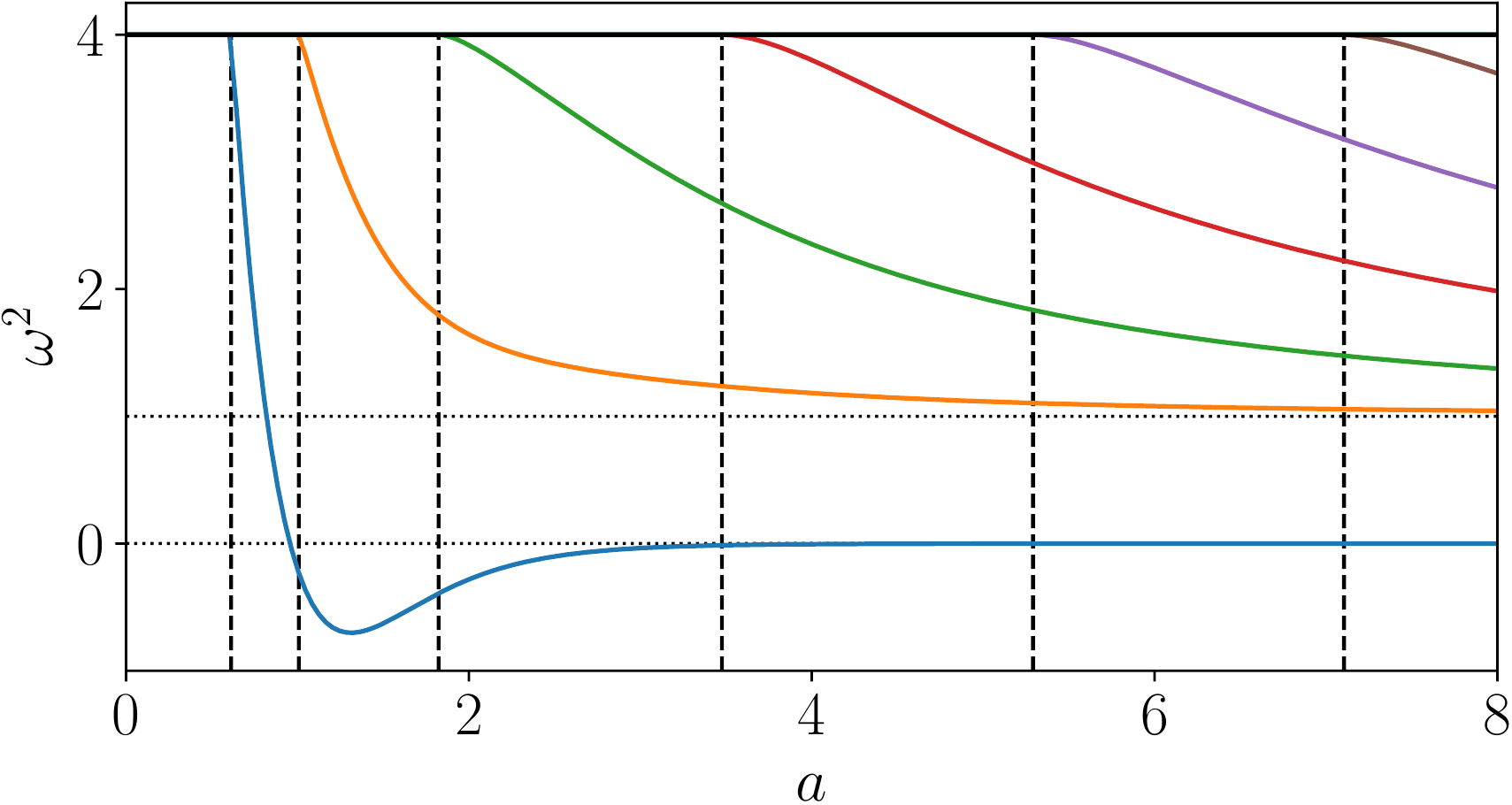}
 \caption{Dependence of the trapped (delocalized) 2-soliton even mode structure on $a$ obtained for the effective potential \ref{AKK-pot-eff} in $\phi^6$ model.}\label{akk_modes_fig}
 \end{figure}

It is known that $\bar{\rm K}{\rm K}$ collisions have a fractal structure in the final state formation \cite{Tr}. Indeed, the incoming antikink and kink may be back scattered, via a sequence of bounces, or may annihilate to the $\phi_v=1$ vacuum, which occurs by the formation of a quasi-periodic kink-antikink bound state, the {\it bion}, which decays to the vacuum by the emission of  radiation. The actual behaviour depends on the initial velocity of the colliding solitons and reveals a fractal-like pattern of bounce windows and bion chimneys, see Fig. \ref{akk_scan_fig}. This is very similar to the famous ${\rm K}\bar{\rm K}$ (and $\bar{\rm K}{\rm K}$) scattering in the $\phi^4$ model. There is, however, one important difference. While solitons in $\phi^4$ theory have a well defined soliton confined DoF, i.e.\ a massive normal mode, the so-called {\it shape mode} or the very similar Derrick mode, the situation in the $\phi^6$ model is more subtle. 

 \begin{figure*}
 \includegraphics[width=1.00\columnwidth]{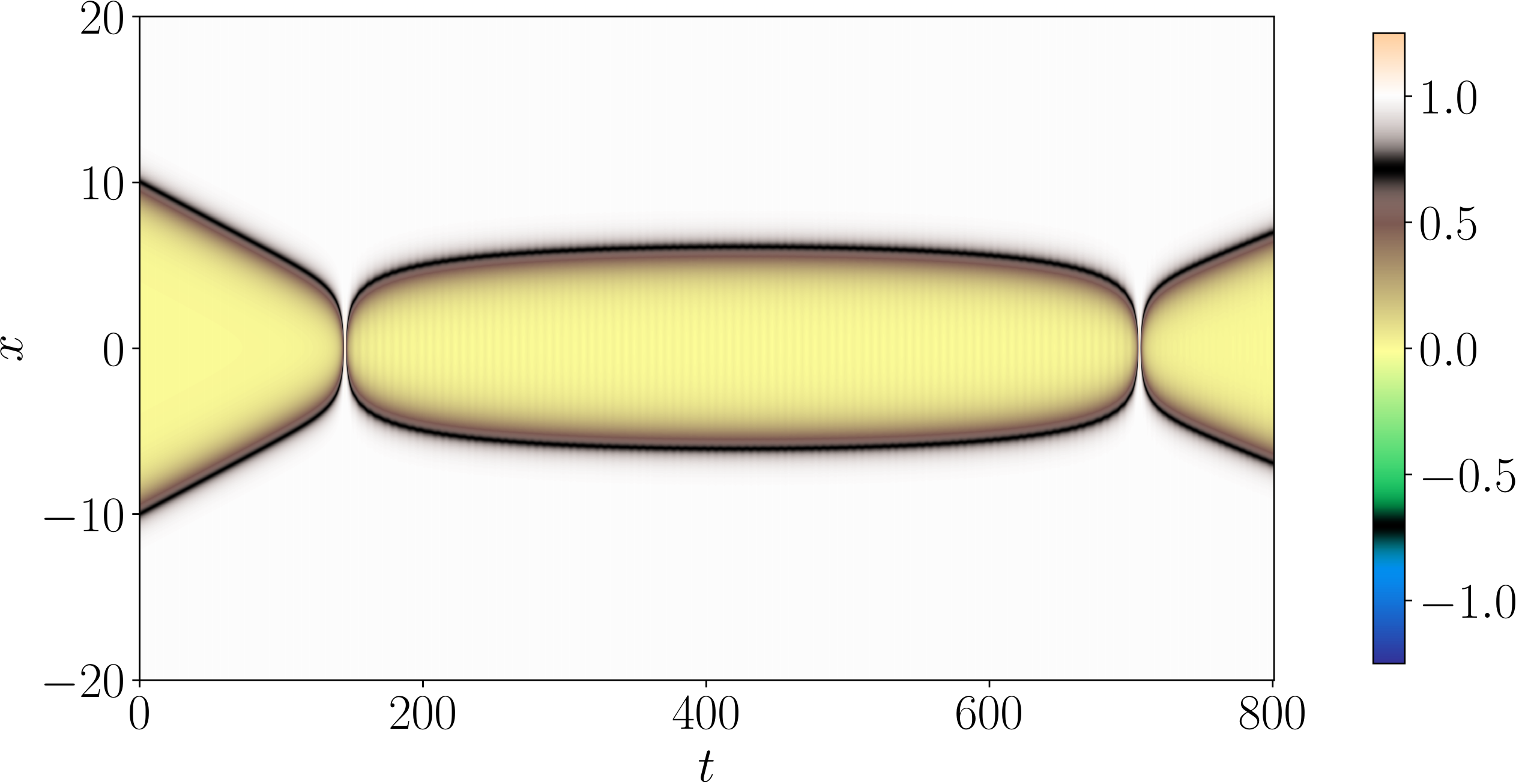}
  \includegraphics[width=1.00\columnwidth]{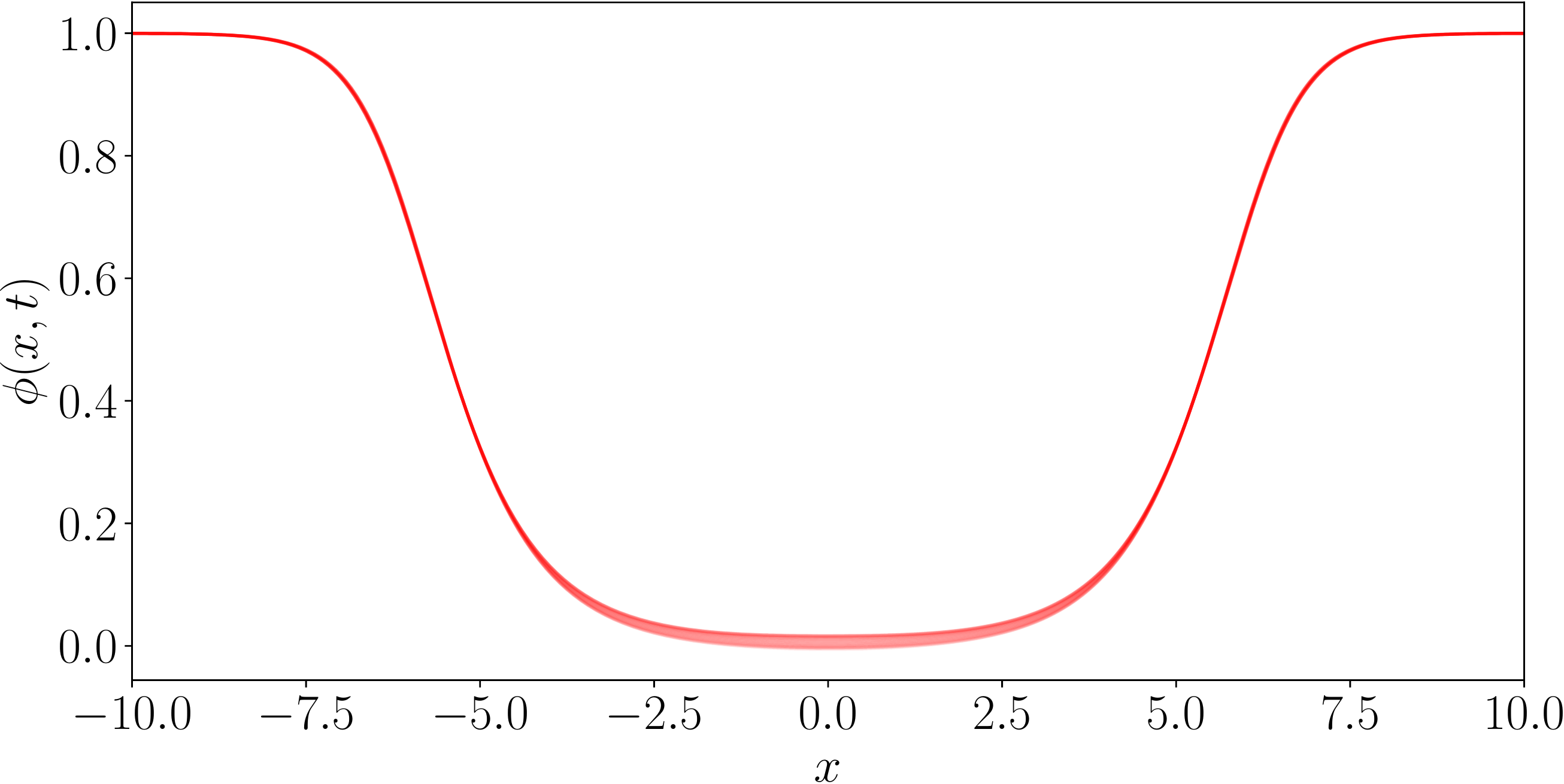}
    \includegraphics[width=1.00\columnwidth]{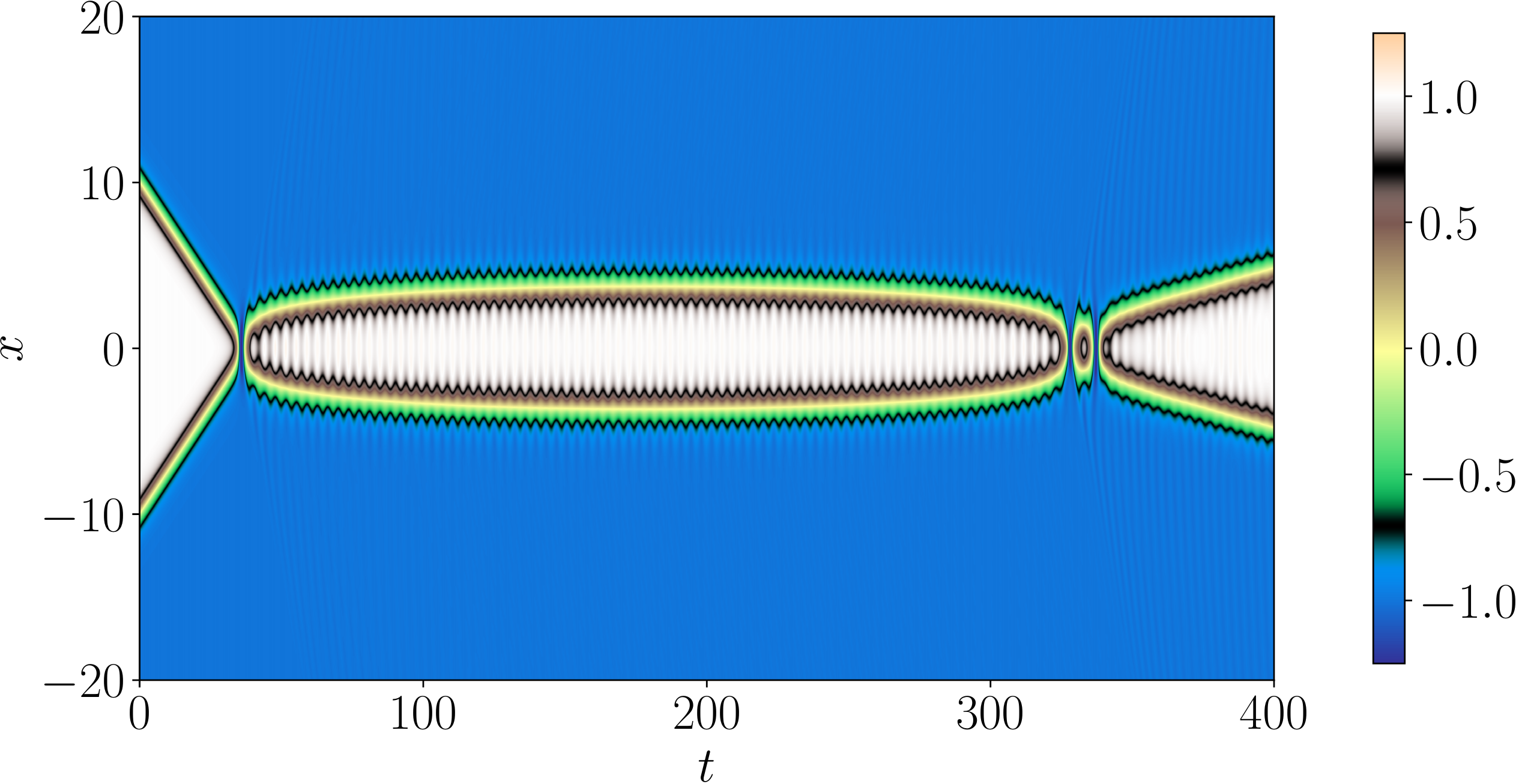}
      \includegraphics[width=1.00\columnwidth]{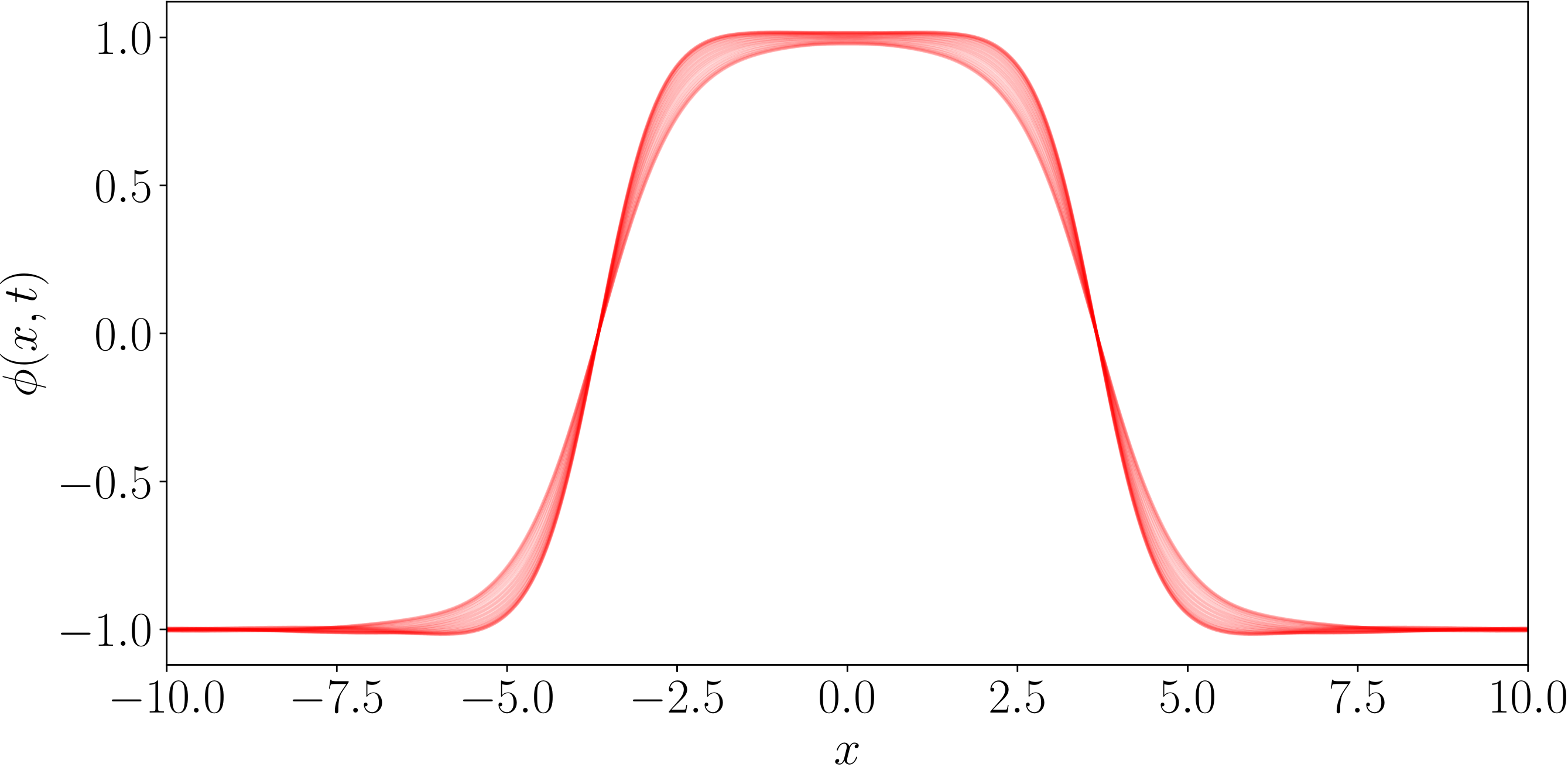}
 \caption{Upper left: an example of a two bounce $\bar{\rm K}{\rm K}$ collision in the $\phi^6$ model with initial velocity of the solitons $v_{in}=0.04542$. Upper right: plot of several profiles for $t\in [422,427]$.  Lower left: an example of two bounce ${\rm K}\bar{\rm K}$ collision in $\phi^4$ model with initial velocity of the solitons $v_{in}=0.2599$. Lower right: plot of several profiles for $t\in [197,202]$.}\label{akk_traj:fig}
        \end{figure*}
\begin{figure}
 \includegraphics[width=1.00\columnwidth]{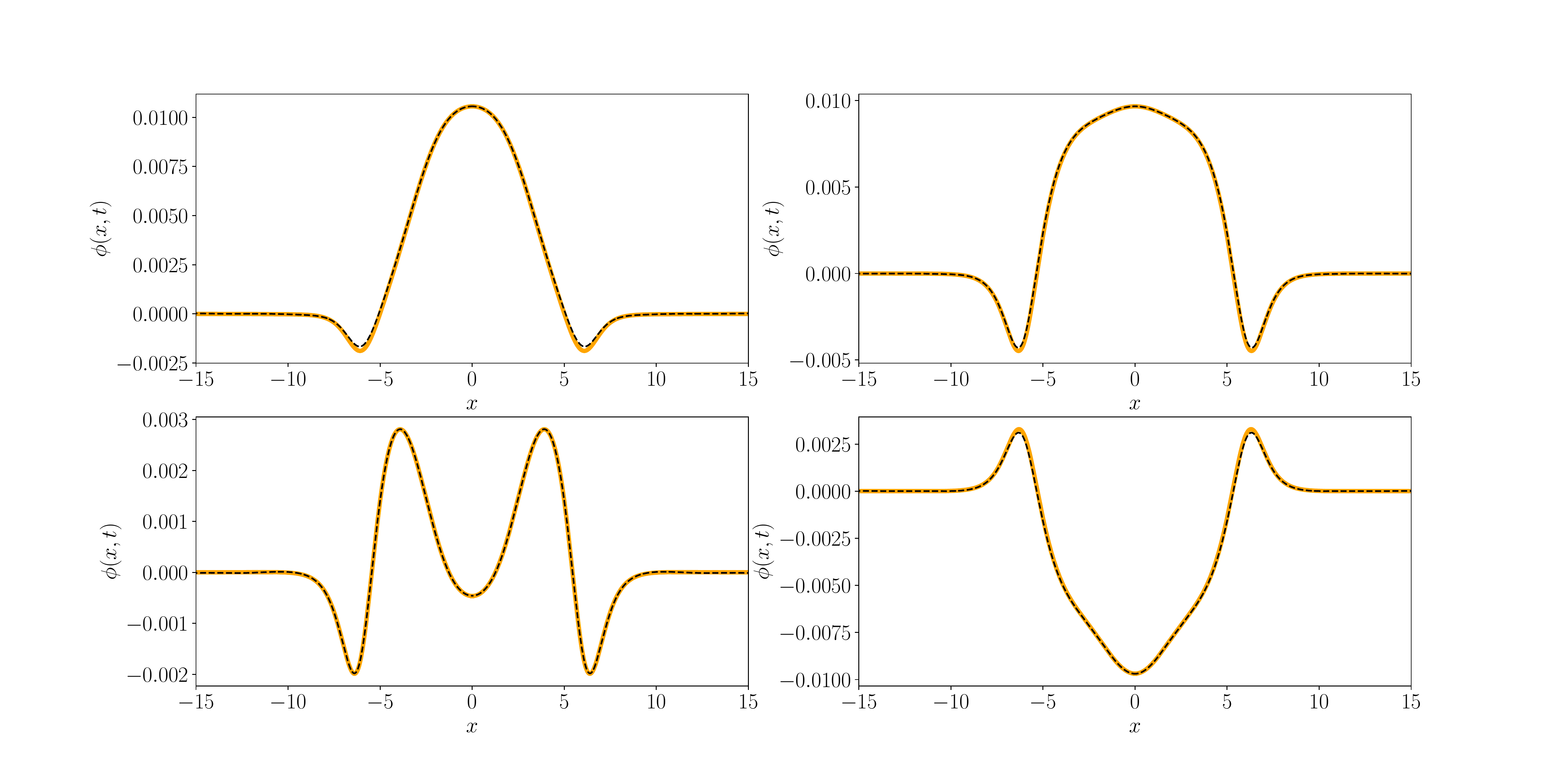}
  \caption{Field profile with subtracted static configuration (\ref{AKK-pair}) for $\bar{\rm K}{\rm K}$ solution in $\phi^6$ model with $v_{in}=0.04542$ (solid line) and the delocalized mode fit (dashed line). }\label{akk_modes_fit:fig}
  \includegraphics[width=1.00\columnwidth]{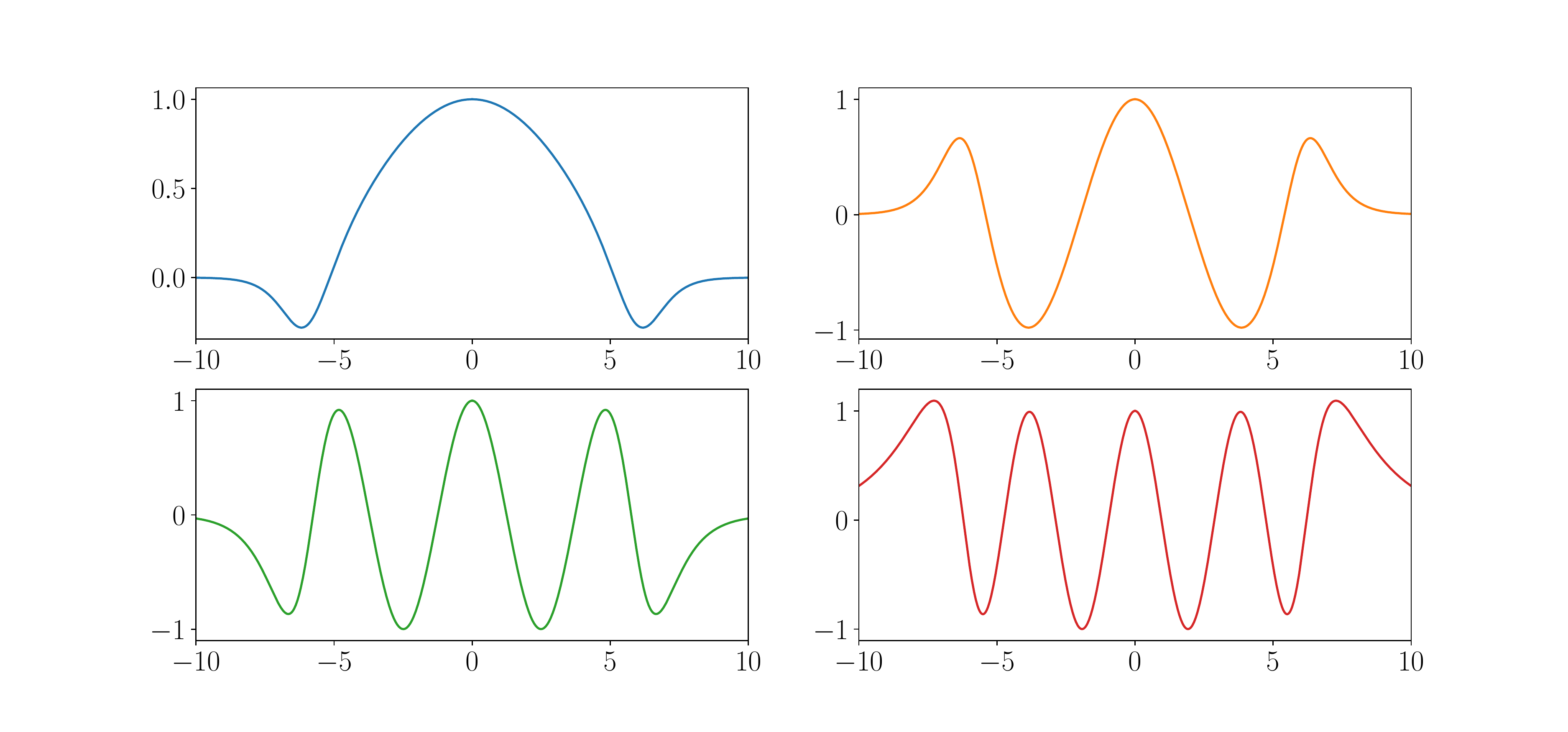}
 \caption{The lowest four delocalized, massive normal modes arising from (\ref{AKK-pair})  with $a=6(?)$.}\label{akk_modes:fig}
        \end{figure}

\begin{figure}
 \includegraphics[width=1.00\columnwidth]{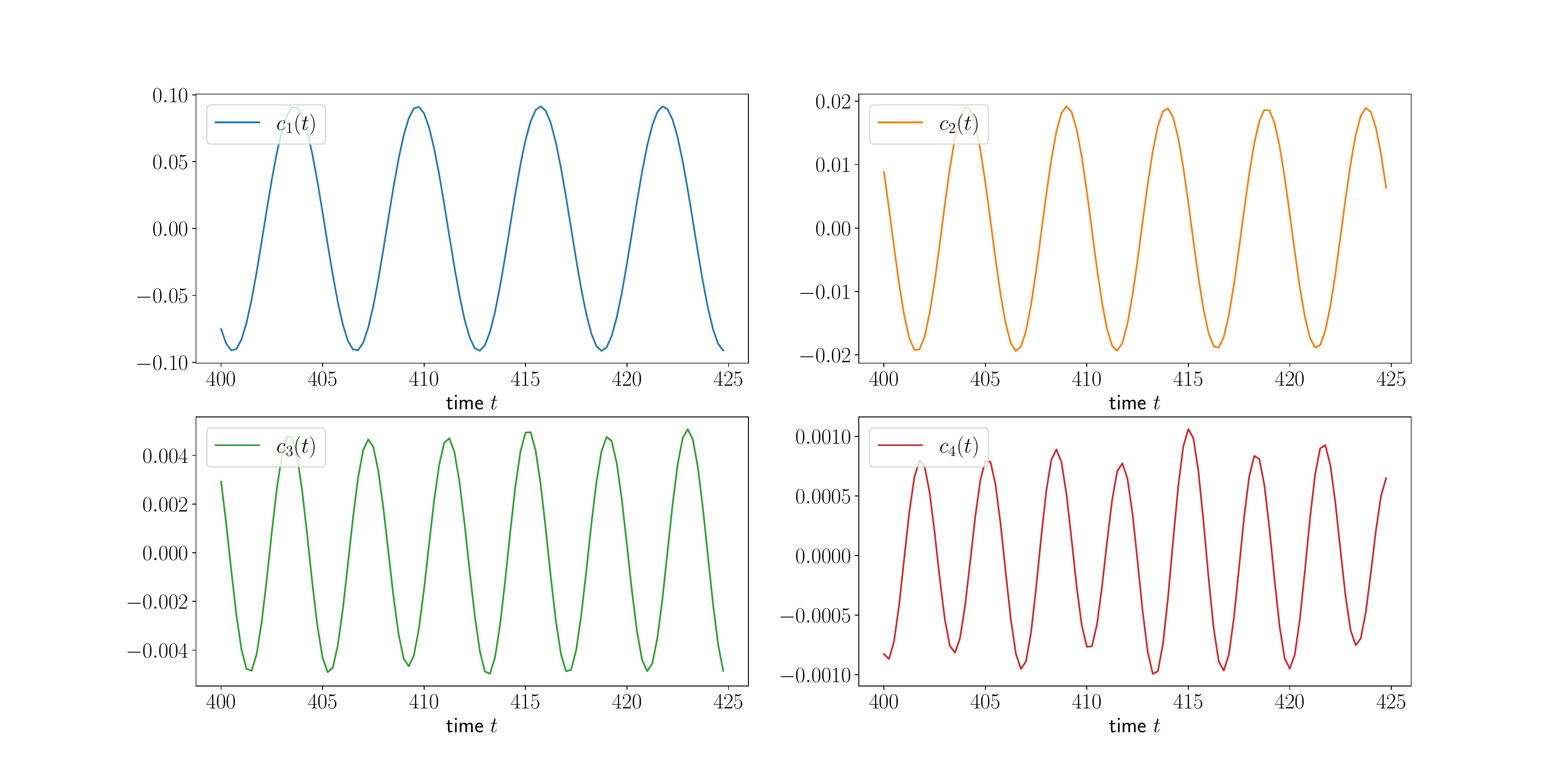}
  \caption{Time dependence of the amplitudes of the first four delocalized modes found in the decomposition of the $\bar{\rm K}{\rm K}$ field profiles for $v_{in}=0.04542$.}\label{modes_time:fig}
     \end{figure}

As we have already noted, a single (anti)kink in $\phi^6$ theory does not possess any massive bound modes which could enter into the resonant energy transfer mechanism and explain the fractal structure. On the other hand, there are soliton confined Derrick modes playing the role of internal DoF. In addition, there are delocalized, two-soliton modes. It was observed in \cite{Tr}, that the naive $\bar{\rm K}{\rm K}$ sum
\be
\Phi_{\bar{K}K}(x;a)=\Phi_{\bar{K}} (x;-a)+\Phi_{K} (x;a), \label{AKK-pair}
\ee 
gives rise to a two-soliton effective potential which hosts trapped or delocalized modes $\eta_i^{\bar{K}K}(x;a)$.  Their form, frequencies and even number depends on the distance between the colliding solitons, see Fig. \ref{akk_modes_fig}. Indeed, there is a zero mode $\eta^{\bar{K}K}_0$ (which corresponds to a simultaneous translation of the $\bar{\rm K}{\rm K}$ pair), an unstable mode $\eta^{\bar{K}K}_{-1}$  (which reflects the fact that the naive superposition is not a static solution of the model) and, finally, there are massive modes $\eta^{\bar{K}K}_i, i\geq 1$,  whose number grows with the intersoliton distance. These two-soliton, trapped modes were used to explain the resonant phenomena in the $\bar{\rm K}{\rm K}$ collision in the $\phi^6$ model \cite{Tr}. 

Here we present further convincing arguments that this is indeed the case - the fractal structure in the final state formation is triggered by the delocalized modes. In addition, we  find  evidence that the Derrick modes have some importance. 

Let us consider a $\bar{\rm K}{\rm K}$ pair colliding with initial velocity $v_{in}=0.04542$. This is an example of a two bounce solution with a rather long lasting bounce window, see Fig. \ref{akk_traj:fig}, upper left panel. There are two important distinct regimes: {\it (i)} the formation of a quasi-stationary state where the positions (centers of mass) of the kink and antikink do not change, and {\it (ii)} the transition through the vacuum, i.e., the annihilation moment. 

As far as the quasi-stationary state is considered, the first  observation is that the change of the field profiles occurs mainly in the region between the solitons. This is shown in Fig. \ref{akk_traj:fig}, upper right panel, where we plot various field profiles for $t \in [422, 427]$. As said, the antikink and kink practically do not change their positions, staying all the time at $a=a_{\rm qs}\approx \mp 6$, while the field {\it between them} fluctuates. This can be contrasted with a two bounce ${\rm K}\bar{\rm K}$ solution in the $\phi^4$ model, see Fig. \ref{akk_traj:fig}, lower panels, where the field fluctuates mainly around the colliding solitons, due to excitation of the shape (or Derrick) mode. Undoubtedly, in the $\phi^6$ model the delocalized excitations seem to govern the dynamics.

This is particularly   visible if we consider the field profiles where we subtract the $\bar{\rm K}{\rm K}$ pair configuration, (\ref{AKK-pair}) with $a=a_{\rm st} =6$ %{YS\color{blue}~ Comment on the choice of this value}, 
see Fig. \ref{akk_modes_fit:fig}, solid line. Here, the profiles are taken at $t=400,401,402$ and $403$. Indeed, the deformation extends over the whole intersoliton region and concentrates in the center. This deformation is fully described by a superposition of the first few two-soliton modes with positive frequency, $\eta^{\bar{K}K}_i$ where $i\geq 1$. As the soliton separation remains basically constant at $a=6$, we use this value to obtain the lowest four normal modes $\eta^{\bar{K}K}_i$, $i=\{1,2,3,4\}$, see Fig. \ref{akk_modes:fig}. We found that the actual field profiles can be decomposed into these first four positive two-soliton modes with a high precision, see, Fig.  \ref{akk_modes_fit:fig}, dashed line. For example, for $t=400$ the corresponding amplitudes are: $A_1= -0.07519, A_2= 0.00881, A_3=0.00292, A_4= -0.00083$. Note that the amplitude of the first delocalized mode $\eta^{\bar{K}K}_1$ is approximately an order of magnitude bigger than the amplitude of the next mode. 
     
The same analysis can be repeated for any $t \in [400,425]$, which still corresponds to the regime when the antikink and kink positions do not change. The resulting time dependence of the amplitudes of the modes is shown in Fig. \ref{modes_time:fig}. Again, the amplitudes of the modes decreases quickly with the mode number. After fitting the periodic functions $A_i^{0} \cos(\omega_i t+\delta)$, we found that the fitted frequencies very well correspond to the frequencies of the modes emerging from the small perturbation around the $\bar{\rm K}{\rm K}$ superposition (\ref{AKK-pair}) with $a=6$. The fitted values are, respectively, $\omega_1=1.0377, \omega_2=1.2756, 
\omega_3=1.5999, \omega_4= 1.9046$ while the frequencies derived in the linear problem are $\omega_1=1.0376, \omega_2=1.2795, 
\omega_3=1.6083, \omega_4= 1.9191$. The agreement is spectacular.  

This fully confirms that the two-soliton (delocalized) modes trigger  the resonant mechanism and, therefore, are responsible for the appearance of the fractal structure. It also firmly shows that these two-soliton modes should be included in any restricted set of configurations, which would be considered as the starting point for a CCM.

 \begin{figure}
 \includegraphics[width=1.00\columnwidth]{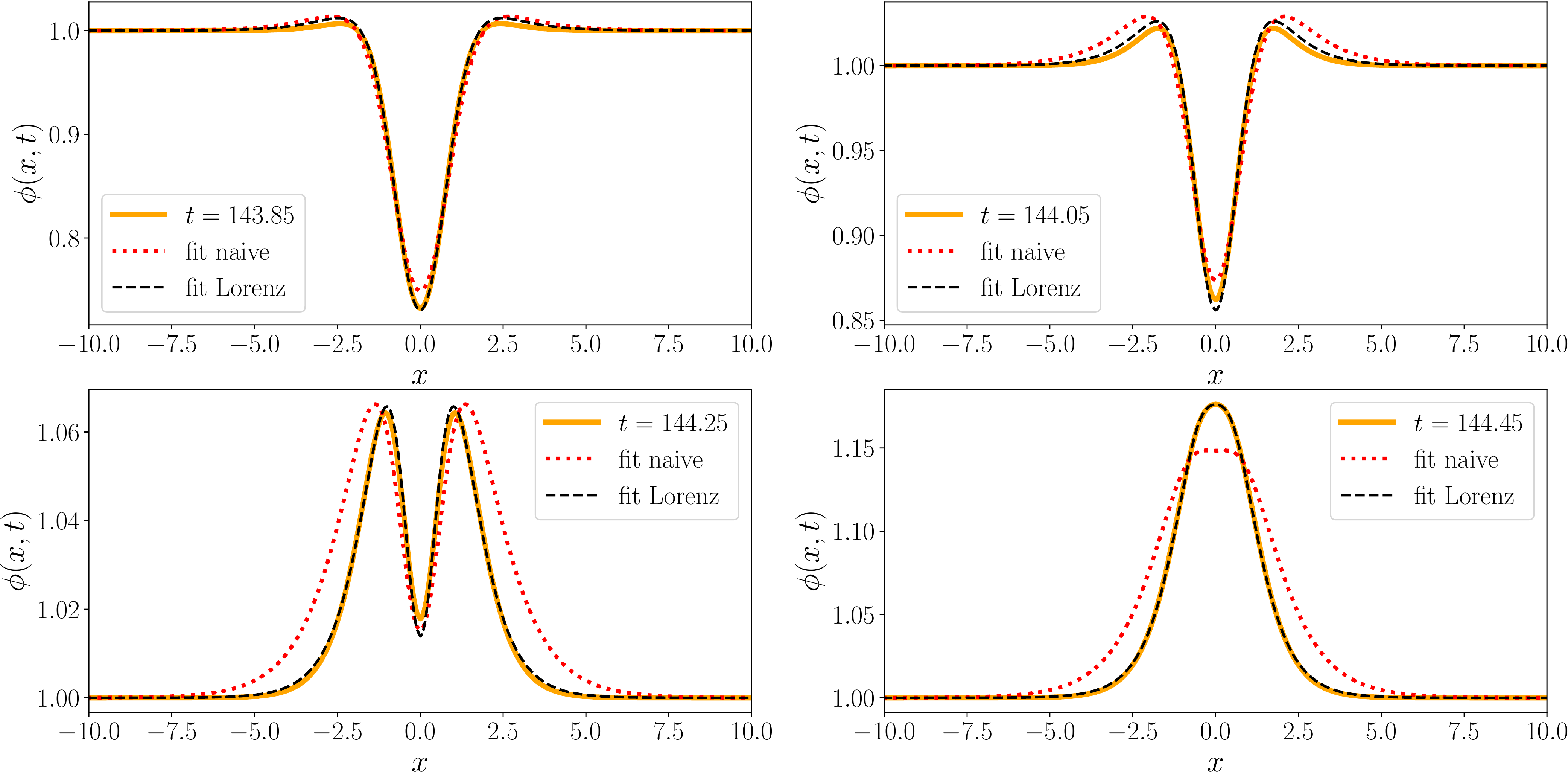}
  \caption{A $\bar{\rm K}{\rm K}$ collision with $v_{in}=0.04542$ at a time close to passing the vacuum. Comparison of field profiles (solid line), naive superposition (\ref{AKK-pair}) (dotted line) and naive superposition with Lorentz factor (dashed line).}\label{lorentz:fig}
     \end{figure}

Now we move to the second regime, that is the annihilation moment.  Although the velocities relevant for two and higher bounce windows as well as for bion chimneys in the $\bar{\rm K}{\rm K}$ scattering in $\phi^6$ model are definitely non-relativistic, with the critical velocity being approximately $v_c\approx 0.048$, passing through the vacuum is a rather violent, relativistic process. To see this, we again consider the $\bar{\rm K}{\rm K}$ solution with $v_{in}=0.04542$ and plot the field profiles at $t \approx 144$, Fig. \ref{lorentz:fig} solid line. These profiles can be approximated by the naive superposition expression (\ref{AKK-pair}) with a moderate success, Fig. \ref{lorentz:fig} dotted line. The fit is significantly improved if we take into account the Lorentz contraction, Fig. \ref{lorentz:fig} dashed line. Here the Lorentz $\gamma$ factor grows from the initial non-relativistic value $\gamma_{in}\approx1.001$ up to $\gamma=1.5$, which corresponds to the highly relativistic velocity $v\approx 0.75$. This suggests a relativistic nature of the field when it crosses the vacuum, even though the solitons start their evolution with totally non-relativistic velocities.

The highly relativistic property of the field in the moment of annihilation can be important for the construction of the correct restricted set of configurations capturing the $\bar{\rm K}{\rm K}$ (or ${\rm K}\bar{\rm K}$) dynamics. The Lorentz contraction of a soliton can be taken into account in terms of the (perturbative) relativistic moduli space by including the  Derrick modes \cite{Rice, AMORW}. Therefore, their amplitudes may play an important role in the construction of the correct CCM, in addition to the amplitudes of the delocalized modes. 

It is instructive to compare this annihilation process with the ${\rm K}\bar{\rm K}$ solution in the sine-Gordon theory, which is an example of a the very special {\it integrable} model, for which 
\be
U_{sG}=1-\cos \phi.
\ee
Here, the ${\rm K}\bar{\rm K}$ scattering solution is exactly known and reads
\be
\Phi_{KAK}=4\arctan \left(\frac{\sinh (\gamma v t)}{v \cosh (\gamma x)} \right).
\ee
Importantly, at any time of the evolution, the ${\rm K}\bar{\rm K}$ solution can be written as a sum of a single kink and antikink at certain positions $\mp a$
\bea
\phi_{{\rm K}\bar{\rm K}}=4\arctan  e^{b (x+a)} -4\arctan  e^{b (x-a)}
\eea
where
\be
b=\gamma, \;\;\; a=\frac{1}{\gamma} \mbox{arsinh} \frac{\sinh (\gamma v t)}{v}.
\ee
As $a$ is the position of the colliding solitons, its time derivative, $\dot{a}$, can be viewed as a velocity of the center of the constituent kink and antikink. Here, 
\be
\dot{a}=\frac{v\cosh(\gamma v t)}{\sqrt{v^2+\sinh^2(\gamma v t)}}
\ee
At $t \to - \infty$, it tends to $v$ which is simply the initial velocity of the infinitely separated, free kink and antikink. However, as $t$ grows, $\dot{a}$ also grows and at $t=0$ it reaches its maximal value  $\dot{a}(t=0)=1$. This is the instant where the solitons annihilate completely, temporarily forming the vacuum. Even though the velocities of the soliton centers tend to 1, one cannot say that the process experiences here a highly relativistic phase. The Lorentz contraction factor $b$ is all the time equal to its initial value $b=\gamma(v)$. Hence, if $v\ll1$ then the non-relativistic approximation, $\gamma \approx 1$ is a valid approximation at any time of the evolution. 

Similar conclusions can be drawn for the sine-Gordon breather. In this case the solution is
\be
b=\sqrt{1-\omega^2}, \;\;\; a=\frac{1}{\sqrt{1-\omega^2}} \mbox{arsinh} \frac{\sqrt{1-\omega^2} \sin (\omega t)}{\omega} \label{breather}
\ee
where $\omega$ is the frequency of the oscillations. As before, $|\dot{a}|\leq1$ with the equality for $t=n\pi$, while the Derrick factor, $b$, does not change. In the non-relativistic approximation $\omega \ll 1$ and once again $b \approx 1$, see e.g., \cite{AMORW}. 

Thus, for the sine-Gordon model we can conclude that an initially non-relativistic kink and antikink remain non-relativistic during the whole evolution.

From the above analysis of the field profiles the following qualitative picture of the $\bar{\rm K}{\rm K}$ scattering in the $\phi^6$ model (in the interesting fractal regime) emerges. Initially, we have an antikink and a kink approaching each other with a small velocity. The profiles are just the naive superposition of the single soliton solutions (\ref{AKK-pair}) and no internal DoF are excited. As the solitons come closer, the profiles require a growing Lorentz contraction factor while simultaneously some additional deformation on the antikink and kink shows up. This deformation spreads quickly over the whole intersoliton region as the solitons approach each other, which eventually leads to the excitation of the lowest delocalized two-soliton modes. Now, the process is very rapid with a highly relativistic $\gamma$. This corresponds to a significant excitation of Derrick modes. A further effect of this is that for even smaller antiki-kink separation, i.e., smaller $a$, for which the positive frequency delocalized modes formally ceases to exist, the profiles still carry such modes. An identical  effect occurs in the ${\rm K}\bar{\rm K}$ collision in the $\phi^4$ model, where the shape mode is not an excitation of the vacuum and, therefore, formally disappears during the annihilation. Apparently, because of the rapidity of the process, the modes are {\it frozen} while the field passes through the vacuum. After crossing the vacuum, the delocalized modes are more and more excited, while the excitation of Derick modes decreases. Thus, the $\gamma$ factor approaches again $1$ and finally, if $v_{in}$ is suitably chosen, the quasi-stationary state is formed. 

In the next section we will analyze several CCM models based on restricted sets of configurations which follow from this general picture. 

%Furthermore, during the annihilation moment, there is no delocalized mode component in the profile. Of course this is expected because in this case, which corresponds to a small value of the modulus $a$, there are no any delocalized modes in the linear perturbation of the naive superposition. Since the process of the crossing of the vacuum is quite rapid, the fact that the delocalized modes at some point (at some intersoliton distance) ceases to exist probably does not affect the dynamics. We will see in sec. \ref{sec:imp} that for slow processes a disappearance of a bound mode (strictly speaking the fact that its frequency equals the mass threshold) may have significant impact on soliton dynamics. Here the energy stored in 

%%%%%%%%%%%%%%%%%%%%%%%%%%%%%
\section{Collective coordinate models}
%%%%%%%%%%%%%%%%%%%%%%%%%%%%%

%%%%%%%%%%%%%%%%%%%%%%%%%%%%%
\subsection{Relativistic CCM with soliton confined modes}
%%%%%%%%%%%%%%%%%%%%%%%%%%%%%
First of all, the traditional way of constructing a CCM, which takes into  account only the zero and shape modes of single solitons, fails completely. As we said, (anti)kinks in $\phi^6$ theory do not host any massive bound modes and, therefore, one is left only with the naive antikink-kink superposition (\ref{AKK-pair}), that is, with the modulus $a$. This is too simple a set of configurations to describe bounces. All solutions of the corresponding CCM with energy bigger than twice the soliton mass are simply one bounce solutions.   

To improve the description, we apply the (perturbative) relativistic moduli space approach for the construction of a (perturbative) relativistic CCM  ((p)RCCM).  This approach in its perturbative version provides an arbitrary large number of coordinates \cite{AMORW}. However, in the simplest set-up, the moduli correspond to single localized soliton excitations and, therefore, cannot be fully suitable for the $\bar{\rm K}{\rm K}$ collisions in the $\phi^6$ model. As a result, the predictions of this minimal version of the (p)RCCM, that is, a version without any addition of the two-soliton delocalized modes, do not agree with the full field theory computations. This is, of course, not surprising in the view of the previous section. We underline that this is not a failure of the perturbative relativistic moduli space framework but rather it reflects a problem of the construction of the two-soliton restricted set of configurations as a naive sum of two single soliton sets, as we explained above. 

Let us start with the relativistic moduli space approach applied to the single kink sector, where the BPS kink static solutions, parametrized by the position $a$, are extended to include the scaling deformation $b$. This results in the following restricted set of configurations 
\be
\Phi_{K} (x;a,b) =\sqrt{\frac{1+\tanh b(x-a)}{2}}
\ee
which give rise to a RCCM possessing a stationary solution equal to a Lorentz contracted kink
\be
\dot{a}=v=\mbox{const}, \;\;\; b=\gamma=\frac{1}{\sqrt{1-v^2}}
\ee
where $v$ is the velocity of a free kink, while $\gamma$ is the Lorentz factor \cite{Rice}. Note that $b \in \mathbb{R}_+$. The same holds for the antikink as well as the mirror solitons. Hence, the inclusion of the new modulus, $b$, leads to a relativistic motion of the kink. In other words, the Lorentz invariance of the original field theory is realized at the level of the CCM.  

Now, we construct the restricted set of configuration for the AKK scattering as a naive sum of two single soliton sets
\bea \label{set-AKK}
&& \mathcal{M}_{\bar{K}K}[a,b]=\{ \Phi_{\bar{K}K}(x;a,b)\}=\nonumber \\
&& \left\{\sqrt{\frac{1-\tanh b(x+a)}{2}}+\sqrt{\frac{1+\tanh b(x-a)}{2}} \right\}
\eea
For $a\to \infty$ we obtain the antikink and kink located at $x=-\infty$ and $x=\infty$ respectively. As $a$ decreases, the solitons come closer and then pass through each other for $a<0$ forming a positive bump whose maximum can never cross $\phi=2$. These configurations provide a two dimensional CCM (\ref{eff-lag})-(\ref{potential}). 

Let us now focus on the moduli space metric resulting from the restricted set of configurations (\ref{set-AKK}). The main observation is that, contrary to the kink-antikink collision in $\phi^4$ theory, the relativistic moduli space metric has no singularity for any finite $a$ as there is no null vector issue. This follows from the fact that the field configurations (\ref{set-AKK}) never equal the vacuum solution $\phi\equiv 1$. Indeed, for any $a$, $\Phi_{\bar{K}K}$, never is a constant field configuration. Specifically, at $a=0$, which is a singular point in the $\phi^4$ case (corresponding to the situation when colliding solitons are on top of each other), the restricted fields have the following expansions
 \bea
 &&\Phi_{\bar{K}K}(x;a,b) = \sqrt{\frac{1-\tanh bx}{2}} +\sqrt{\frac{1+\tanh bx}{2}}  \nonumber \\
 && \left( \frac{1-\tanh^2bx}{\sqrt{1-\tanh bx} }+\frac{1-\tanh^2bx}{\sqrt{1+\tanh bx}} \right) \frac{ab}{2\sqrt{2}} + o(a)
 \eea
 Hence, $\partial_a \Phi_{\bar{K}K}$ is not identically zero for $a=0$, leading to non-zero metric components. There is still a singularity at $b=0$. However, as we will discuss below, this point cannot be attained by a finite energy configuration and, therefore, is excluded from the dynamics. Hence, the RCCM is {\it globally} well defined for the scattering case.  

Finally, we consider the effective potential $V$ (\ref{potential}).  In the limit $b \to 0$, where the metric has the singularity, the restricted configurations tend to a constant which is not a vacuum. Namely, $\Phi_{\bar{K}K }(x,a,  b=0)=\sqrt{2}$. Obviously, this corresponds to infinite potential energy. This means that the singularity of the metric at $b=0$ is never accessible for a finite energy configuration. 

To conclude, the relativistic moduli space modeling the antikink-kink collisions in $\phi^6$ theory is a smooth two-dimensional manifold without any singularities. Therefore, it leads to a globally well-defined dynamical RCCM. Owing to this property we can study the collisions already in the RCCM. This is a nice feature which is not shared by the RCCM for scatterings of symmetric kinks, as in the $\phi^4$ model, where a regularizing scheme had to be introduced. 

\begin{figure}
 \includegraphics[width=1.00\columnwidth]{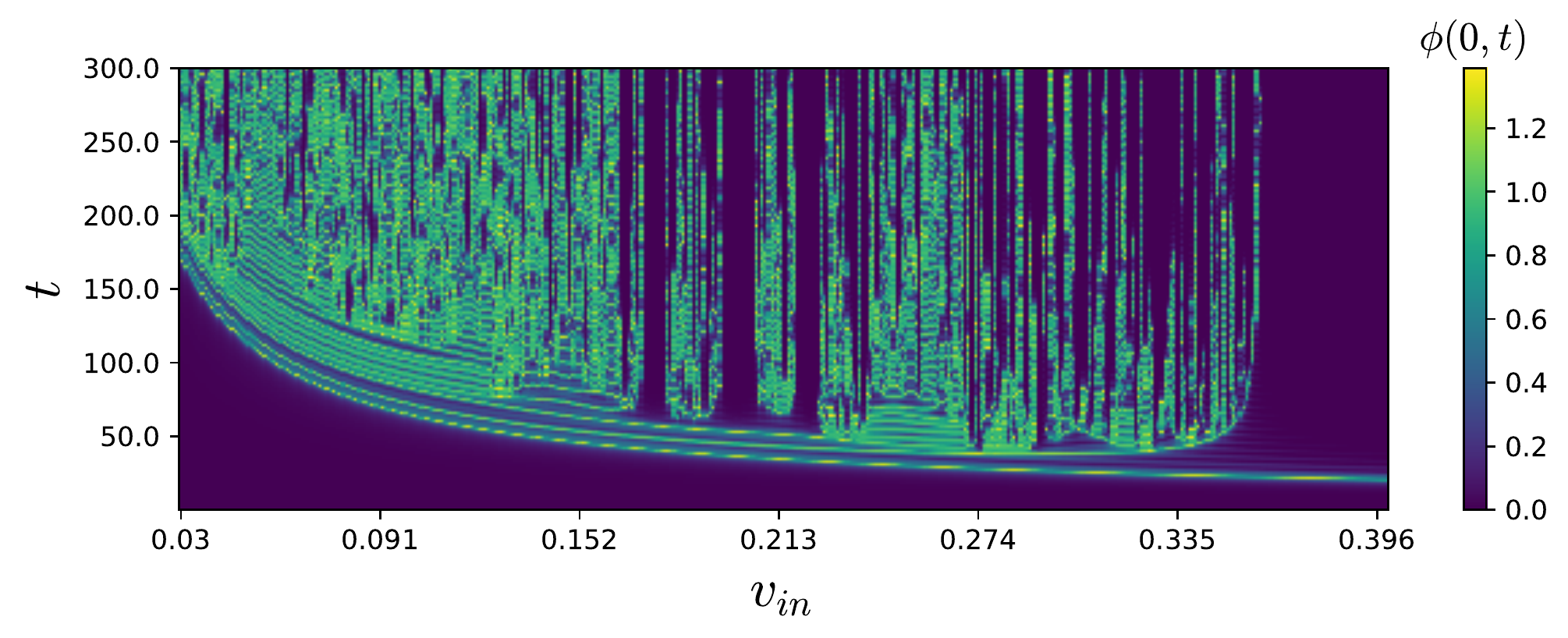}
 \includegraphics[width=1.00\columnwidth]{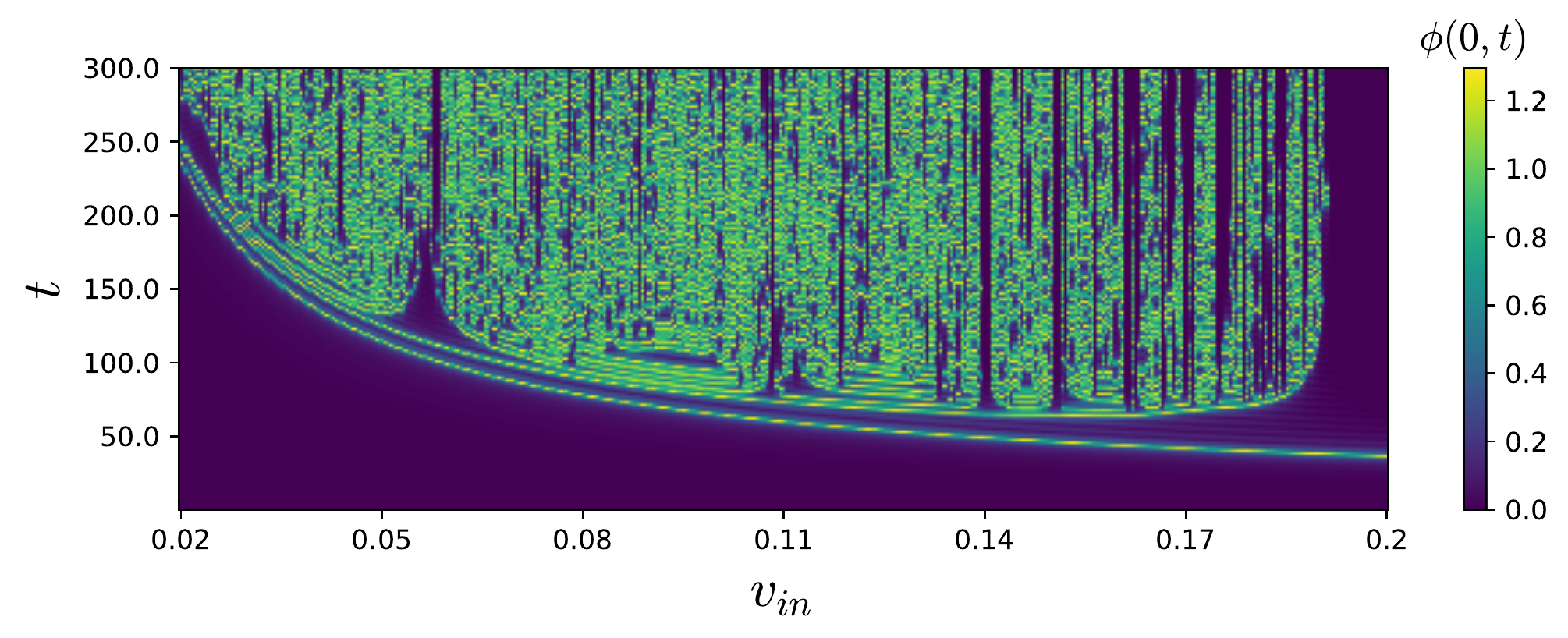}
  \caption{Time dependence of the value of the field at the origin, $\phi(x=0,t)$ for various initial velocities $v_{in}$ in CCMs for the $\bar{\rm K}{\rm K}$ collision in $\phi^6$ model. Upper: RCCM  with moduli $(a,b)$. Lower: pRCCM with moduli $(a,c)$. }\label{RCCM:fig}
     \end{figure}

In Fig. \ref{RCCM:fig}, upper panel, we show the scan of the evolution obtained in the corresponding RCCM  (\ref{eff-lag})-(\ref{potential}). The initial conditions are
\be
a(0)=a_0, \;\;\; \dot{a}(0)=v_{in}, \;\;\; b(0)=\gamma_{in}, \;\;\; \dot{b}(0)=0
\ee
where the initial separation $2a_0=24$. We show the value of the field as a function of time for different values of the initial velocities. Qualitatively, the effective model possesses desirable properties. It predicts two, three and higher bounce windows immersed between bion chimneys. However, the critical velocity, which separates the fractal regime and one bounce scattering is much higher than in the original theory. Instead of $v_c\approx 0.048$ it is $v_c\approx 0.37$. There are two related reasons for this failure. On the one hand, the restricted set of configurations does not approximate the true profiles too well. For example, $\Phi_{\bar{K}K}(x;a,b)$ is always bounded between $0 <\phi(x,t)<2$. On the other hand, it does not have any wiggles in the central region. This corresponds to the fact that the scale deformation is confined to the solitons. Hence, our RCCM contains only soliton confined DoF.  

This restricted set of configurations can be improved if we consider a variant of the upper construction, known as the perturbative relativistic moduli space. We present it starting with the single soliton sector, $\Phi_K(x; a,b)$. Now we expand the scaling deformation of the (anti)kink, $b=1+c$, assuming that $c$ is a small parameter. Here we consider only the first term in the expansion
\bea
& \Phi_K(x; a, c) & \approx \sqrt{\frac{1+\tanh(x-a)}{2}} \nonumber  \\
&&+\frac{(x-a)c}{2\sqrt{2}\cosh^2(x-a)\sqrt{1+\tanh(x-a)}}  \label{rest-pert}
\eea

In principle, we can keep an arbitrary number of the terms in the expansion, which is equivalent to the fact that we take into account further Derrick modes. The crucial idea is to replace each power of the expansion parameter, $c^n$, by a new, independent collective coordinate (amplitude) $C_n$, of the pertinent Derrick mode. Inserting the perturbative expansion of the restricted configurations (\ref{rest-pert}), we obtain a pRCCM with the following metric functions
\bea
g_{aa}^K&=& \frac{1}{96} (24+24c+(6+\pi^2)c^2)\\
g_{cc}^K&=& \frac{\pi^2}{48}, \;\;\; 
g_{ac}^K= \frac{1}{8}
\eea
and the effective potential
\bea
V_K&=&\frac{1}{4} + \frac{c^2}{8} + \frac{c^3}{96}(\pi^2-6) + \frac{c^4}{19 200} (570-275 \pi^2 +21\pi^4) \nonumber \\
&+& \frac{c^5}{11 520} (-90+75\pi^2-7\pi^4) \nonumber \\ 
&+& \frac{c^6}{1024} \left( \frac{6}{7} - \frac{7\pi^4}{20} + \frac{31 \pi^6}{882}  \right).
\eea
In the case of the antikink sector, the moduli space metric differs only in the off-diagonal term, $g^{\bar{K}}_{ac}=-1/8$, while the effective potential is exactly the same.

The pRCCM in the kink or antikink sector supports a stationary solution with $\dot{a}=v=\mbox{const}$, $c=\tilde{c}=\mbox{const}$ following an algebraic  equation
\be
\frac{v^2}{2} \frac{dg_{aa}}{dc} = \frac{dV}{dc}
\ee
This solution is an approximation of a Lorentz boosted (anti)kink. 

The construction of the pRCCM describing antikink-kink collisions goes along standard lines. The corresponding restricted set of configurations is built as a simple sum of the single soliton sectors. Hence,
\be
\Phi_{\bar{K} K}= \Phi_{\bar{K}} (x; - a, c) + \Phi_K(x; a, c), \label{conf-pRCCM}
\ee
which leads to a pertinent collective model. Again, one can easily verify that the restricted configurations do not produce any null vector problem and the moduli space metric is well defined everywhere. 

As the initial states in the soliton collisions are a free, boosted kink or antikink, we have to specify the corresponding initial conditions. These are the stationary solutions found in the single soliton pRCCM. Thus,
\be
a(0)=a_0, \;\;\; \dot{a}=v_{in}, \;\;\; c(0)=\tilde{c}, \;\;\; \dot{c}=0
\ee
Here $a_0=12$ is half of the initial separation between the colliding kink and antikink and $v_{in}$ their initial velocity. 

The results are  better than in the RCCM but the critical velocity is still approximately four times bigger than in reality, see Fig, \ref{RCCM:fig}, lower panel. The improvement is probably related to a slightly better set of field configurations whose value is now not bounded to the segment $(0,2)$. Of course, the first Derrick mode is confined to the solitons and therefore cannot alone correctly model the true dynamics. 

We expect that a further improvement can be achieved by considering higher Derrick modes. They are more and more widespread modes which potentially may better approximate the true delocalized two-soliton modes. 

Summarizing the results of this section, we can draw the conclusion that a CCM based entirely on soliton confined DoF cannot successfully describe the $\bar{\rm K}{\rm K}$ dynamics in the $\phi^6$ model. The inclusion of the delocalized, two-soliton modes seems to be inevitable. 

%%%%%%%%%%%%%%%%%%%%%%%%%%%%%
\subsection{CCM with one delocalized mode}
%%%%%%%%%%%%%%%%%%%%%%%%%%%%%

As we have clearly shown above, during the scattering one observes the excitation of the delocalized modes $\eta^i_{\bar{K}K}(x;a)$, which arise in the linear perturbation theory of the kink-antikink state. Therefore, we propose the following restricted set of configurations 
\bea
& &\tilde{\mathcal{M}}_{\bar{K}K}[a;X^1...X^N]= \nonumber\\
&&\{ \Phi_{\bar{K}}(x; -a)+\Phi_{K}(x; a)+ \sum_{i=1}^NX^i  \eta^{\bar{K}K}_i(x,a) \}  .
\eea
We also observed that the first delocalized mode $\eta^{\bar{K}K}_1$ plays the most significant role, as it stores the biggest amount of energy. Thus, it is reasonable to start with as simple a CCM as possible and consider only the first delocalized mode
\be
\tilde{\mathcal{M}}_{\bar{K}K}[a;X^1]= 
\{ \Phi_{\bar{K}}(x; -a)+\Phi_{K}(x; a)+X^1  \eta^{\bar{K}K}_1(x,a) \}. \label{set_aX}
\ee
The mode still has quite an involved dependence on the position $a$ of the colliding solitons. Furthermore, for sufficiently small $a$ it disappears from the spectrum of linear normal modes.  To simplify our considerations, however,  we will not take into account this variation with $a$. Instead, in our simplest CCM we assume that the mode does not change and is equal to the first delocalized mode at a given $a=a_*$. This is a standard assumption for the construction of a CCM for kink collisions referred to as the {\it frozen vibrational moduli space approximation}. For example, it has been successfully applied to ${\rm K}\bar{\rm K}$ scattering in the $\phi^4$ model \cite{MORW}. This defines the final form of the restricted configurations
\be
\tilde{\mathcal{M}}_{\bar{K}K}[a;X^1]= 
\{ \Phi_{\bar{K}}(x; -a)+\Phi_{K}(x; a)+X^1  \eta^{\bar{K}K}_1(x,a=a_*) \}.  \label{set_aX_f}
\ee
The reason why this is an admissible assumption is related to the fact that the process is always rapid (relativistic) when the kinks are very close to each other. This explains why the disappearance of the internal modes (strictly speaking their crossing through the mass threshold) in many models does not seem to affect the ${\rm K}\bar{\rm K}$ scattering. The reason is that the mode ceases to exist approximately in the moment when the field is close to the vacuum value, i.e., when the solitons are almost on top of each other.  However, precisely in the moment of temporary annihilation, the scattering solitons behave very relativistic which translates into a rapid evolution of the field. If the typical time scale of the evolution is smaller than the time scale of the internal mode oscillation, then the mode freezes, preserving its form despite the change of the underlying field. This is very much alike the non-adiabatic evolution of a quantum-mechanical wave function if the quantum Hamiltonian changes very rapidly (instantaneously) from $H(t_0)$ to $H(t_1)$. After the temporary annihilation, when the solitons reappear and the system returns to the adiabatic evolution, the mode is again able to evolve with its soliton(s). 

\begin{figure}
 \includegraphics[width=1.00\columnwidth]{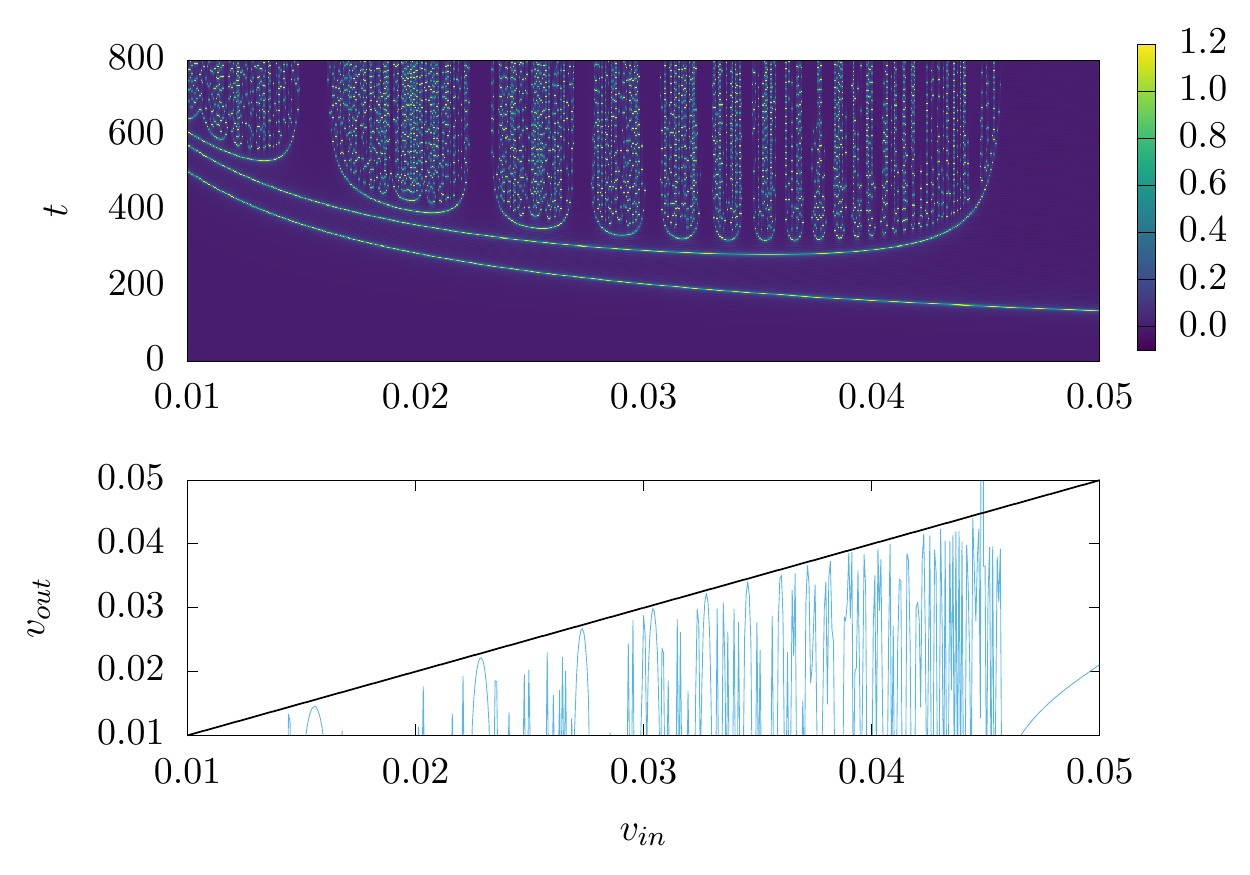}
  \caption{$\bar{\rm K}{\rm K}$ collisions in the $\phi^6$ model in the CCM based on $\tilde{\mathcal{M}}[a,X^1]$:  time dependence of the field at the origin, $\phi(x=0,t)$, and final velocity of the backscattered antikink, $v_{out}$, for various initial velocities~$v_{in}$.}\label{CCM_aX:fig}
     \end{figure}

All this results in the following final form of the restricted set of configurations 
\bea
& &\tilde{\mathcal{M}}_{\bar{K}K}[a;X^1]= \\
&&\left\{\sqrt{\frac{1-\tanh (x+a)}{2}}+\sqrt{\frac{1+\tanh (x-a)}{2}}+ X^1  \eta^{\bar{K}K}_1(x) \right \}, \nonumber
\eea
where 
\bea
\eta^{\bar{K}K}_1(x) &=& \frac{\cos(x\pi/7)}{2} \left( \tanh(x+3) -\tanh(x-3)\right)\nonumber \\
&+&\frac{1}{10} \left( -2+\tanh(x+3) -\tanh(x-3)\right)\times \nonumber \\
 &\times& \left( \frac{1}{\cosh(x+3)}+\frac{1}{\cosh(x+3)} \right) 
\eea
is an approximate analytical form of the first delocalized mode for the particular value $a_*=3$.

This restricted set of configurations defines a two dimensional CCM with a non-singular moduli space and potential for any finite $(a,X^1) \in \mathbb{R}^2$. Hence we get a well defined dynamical system $L(a,X^1)$.  We do not present the explicit formulas for $g_{ij}$ and $V(a,X^1)$ because they are very complicated and because in our numerical scheme we do not use analytical expressions but compute these functions numerically. 

Finally, the initial conditions for the evolution in the CCM are as follows
\be
a(0)=a_0, \;\; \dot{a}(0)=v_{in}, \;\;\; A(0)=0, \;\; \dot{A}(0)=0
\ee
These are natural initial conditions corresponding to non-relativistic kinks with no excited modes. 

In Fig. \ref{CCM_aX:fig} we present our results. In particular, we plot the time dependence of the field at the origin for initial velocity $v_{in} \in [ 0.01, 0.05]$. We found a spectacular {\it quantitative} agreement with the full field theory computation. 

First of all, exactly as in the full field theory, the fractal structure shows up in the non-relativistic range of initial velocities. The critical velocity, above which only one-bounce scattering occurs, is $v_{crit}^{CCM} = 0.0457$ which is exactly equal to the true value $v_{crit}=0.0457$ within our numerical precision.

Secondly, almost all positions of the two-bounce windows, as well as their widths, coincide with the full theory computation. Only the two-bounce windows located around $v_{in}\approx 0.015$ and $v_{in}\approx 0.0275$ correspond to false windows in the full model.  This suggests that the existence of false windows may be related with higher delocalized modes or with radiation. Undoubtedly, our collective model gives a tool in which this question can be analyzed. 

It is important to remark that the results of the CCM depend on the value of the modulus $a=a_*$ at which we freeze the first delocalized mode. The critical velocity grows from $v_{cr}=0.0320$ for $a_*=2.6$ to $v_{cr}=0.0598$ for $a_*=3.5$, see Table 1. Details of the fractal structure also vary with the freezing point. 

\begin{table} 
 \label{tab}
{\scriptsize 
\begin{tabular}{cc} 
\hline\hline
\hspace*{1cm}$a_*$\hspace*{1cm} & \hspace*{1cm} $v_{cr}$ \hspace*{1cm} \\
\hline
2.6 &  0.0320 \\
2.7 &  0.0355\\
2.8 &  0.0391\\
2.9 &  0.0424\\
3.0 &  0.0457\\
3.1 &  0.0487\\
3.2 &  0.0517\\
3.3 &  0.0545\\
3.4 &  0.0572\\
3.5 & 0.0598 \\
\hline\hline
\end{tabular}} 
\caption{Dependence of the value of the critical velocity $v_{cr}$ obtained in the CCM based on  (\ref{set_aX_f}) on the freezing point $a_*$.}
\end{table}

Of course, in a more complete CCM based on the dynamical restricted set of configurations (\ref{set_aX}) one should take into account the fact that the delocalized mode changes during the collision and does not freeze at a given $a$. Here, however, we follow the simplest strategy. Namely, we just select a particular shape of the first delocalized mode, i.e., freeze it at a certain $a_*$. Therefore, this freezing point $a_*$ may be viewed as a free parameter of the simplest CCM. For $a=a_*=3$ we found the most striking agreement. 
This value is "natural" in the sense that quasi-stationary states with a $\bar{\rm K}{\rm K}$ distance of $2a_{\rm qs} \sim 2a_* \sim 6$ form in many bounce windows. As a general rule, the value of $2a_{\rm qs}$ grows with the lifetime of the quasi-stationary state. In section III, e.g., we analyzed in detail the case of a particularly long-lived quasi-stationary state with $a_{\rm qs} =6$. Probably, the correct separation in each bounce window can be reproduced in a more complete CCM provided by (\ref{set_aX}). 

In any case, we must admit that at the moment we do not fully understand why
our simple CCM for $a_* =3$ agrees so spectacularly well with the full field theory for {\em all} initial velocities. Naively, one would probably expect a certain variation of the optimal value for $a_*$ with the initial velocity. A more complete understanding of this issue most likely requires a dynamical treatment of the delocalized modes where $a$ is not frozen but allowed to vary with time.
It must be stressed, however, that such a fully dynamical treatment within a CCM framework is a very challenging problem. This reflects the existence of essential singularities of any moduli space based on delocalized normal modes. Indeed, any of these modes ceases to exist for a sufficiently small antikink-kink distance. One by one they hit the mass threshold and become non-normalizable threshold modes which further change into anti-normal or quasinormal modes. As a consequence, some of the metric components diverge at these points which results in a breakdown of a CCM based only on normal modes \cite{spectral-wall}. 

%%%%%%%%%%%%%%%%%%%%%%%%%%%%%
\section{${\rm K}\bar{\rm K}$ collisions}
%%%%%%%%%%%%%%%%%%%%%%%%%%%%%

\begin{figure}
 \includegraphics[width=1.00\columnwidth]{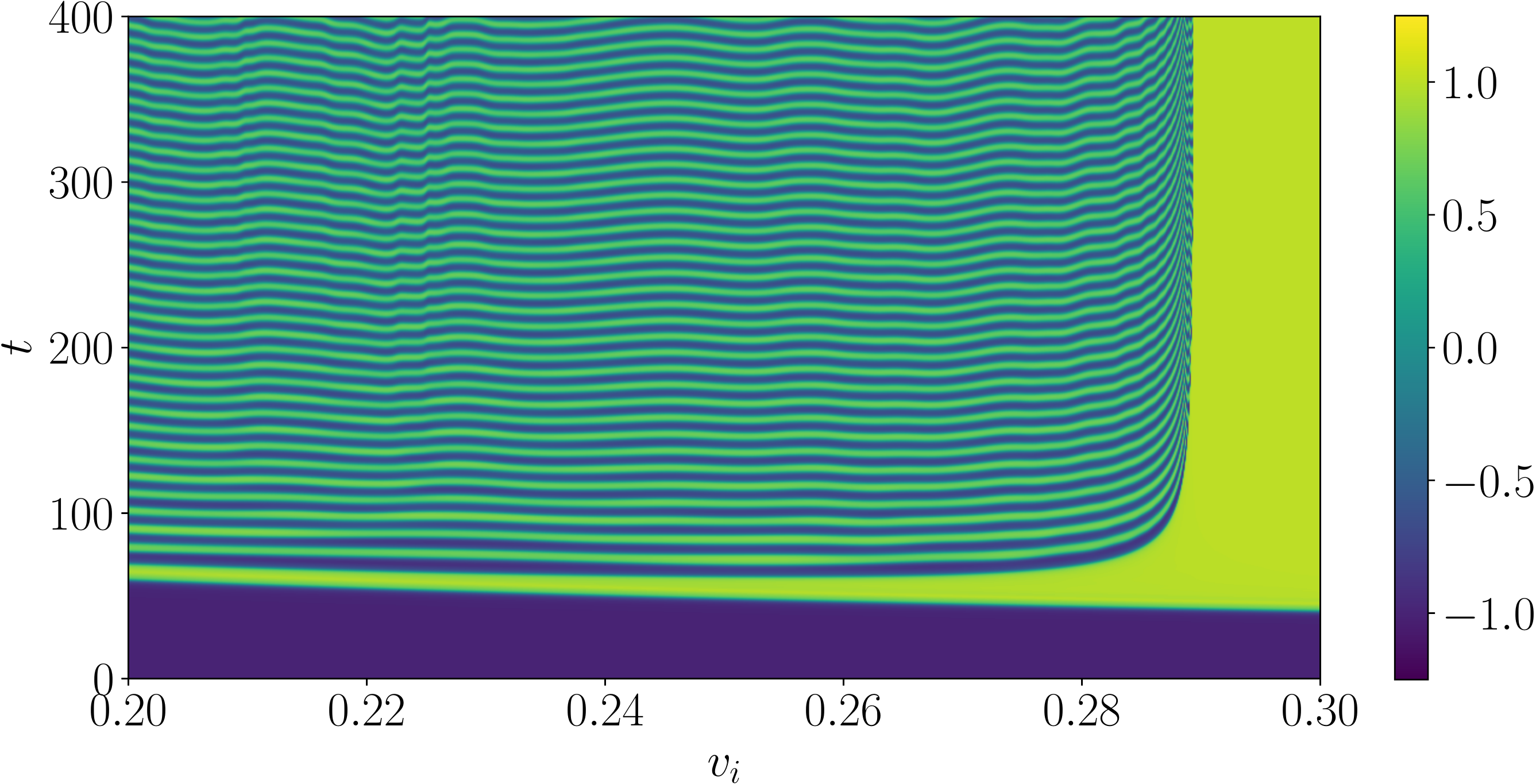}
  \caption{Time dependence of the value of the field at the origin, $\phi(x = 0, t)$, for various initial velocities $v_{in}$ in  ${\rm K}\bar{\rm K}$ collisions in the $\phi^6$ model. }\label{KAK:scan}
     \end{figure}

In this section we turn to the kink-antikink collisions. It is known that in this case there is no fractal structure in the final state formation. Instead, there are two regimes which meet at $v_{cr}\approx 0.289$. For smaller initial velocities, the solitons undergo annihilation via the formation of a quickly oscillating bion/oscillon, while for $v>v_{cr}$ they simply pass through each other and change into their mirror kinks which escape to infinity, see Fig. \ref{KAK:scan}.  Note that the critical velocity is quite big if compared with the $\bar{\rm K}{\rm K}$ case. 

Despite the relatively simple outcome of this scattering, it is surprisingly nontrivial to model it within the collective coordinate framework. 
This originates in the fact that now radiative modes play a more significant role than in  the previous $\bar{\rm K}{\rm K}$ case or in ${\rm K}\bar{\rm K}$ collisions in the $\phi^4$ model. Indeed, there are no single-kink or delocalized kink-antikink normal modes involved here. Therefore, the main channel in which energy escapes from the kinetic DoF of the solitons is radiation. Ideally we would like to have a CCM which includes such radiative DoF. This is a complicated and still not solved issue within the CCM framework. However, as we have noticed a few times, the Derrick modes may capture some properties of radiation, at least at short time scales. The description in terms of these modes should become better as we increase their number. In the subsequent analysis we apply this approach, again in the simplest version, i.e., with only one Derrick mode taken into account. Obviously, this is a very crude approximation which cannot lead to satisfactory results. Nonetheless, some findings are encouraging and give new insights into the understanding of this process. 

\begin{figure}
 \includegraphics[width=1.00\columnwidth]{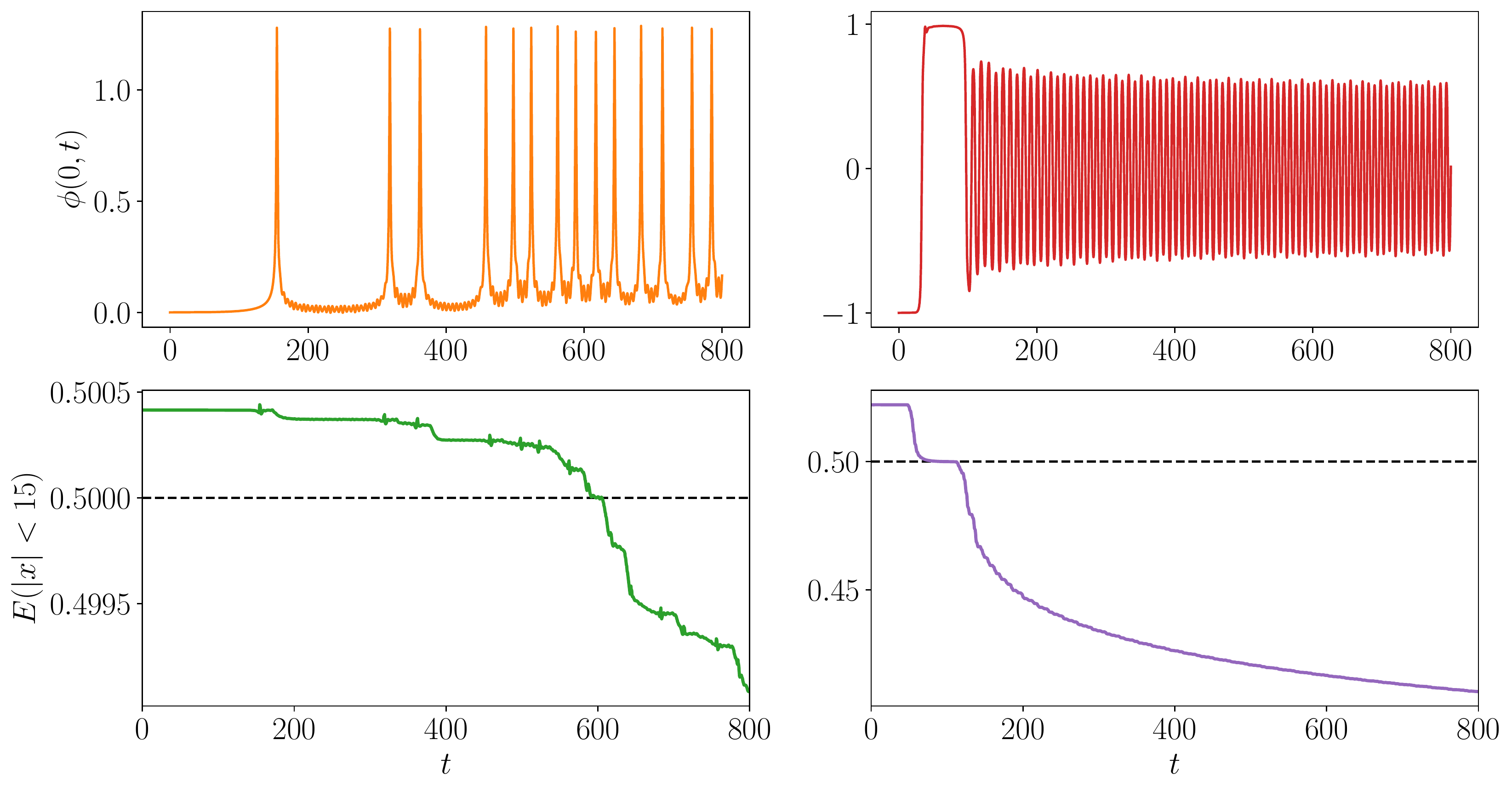}
  \caption{Value of the field at the origin (upper panels) and energy stored in the region $|x|<15$ (lower panels) for a $\bar{\rm K}{\rm K}$ collision with $v_{in}=0.042$ (left panels) and  a ${\rm K}\bar{\rm K}$ collision with $v_{in}=0.288$ (right panels). }\label{KAKvAKK}
     \end{figure}

To see that there is significant radiation in a ${\rm K}\bar{\rm K}$ collision in the $\phi^6$ model we consider a kink and antikink boosted towards each other with initial velocity $v_{in}=0.288$, which is just below the regime where the simple passing-through scenario occurs. We compare it with a $\bar{\rm K}{\rm K}$ collision with $v_{in}=0.042$. In Fig. \ref{KAKvAKK} we plot the time evolution of the field in the origin (upper panels) together with the energy stored in the space region $-15<x<15$ (lower panels). Thus it shows how much of the energy is radiated away from the solitons.  

     
In the left panels we show the $\bar{\rm K}{\rm K}$ collision. We see that after each bounce, where the field passes $+1$ vacuum, the part of the energy is emitted from the solitons. The amount of emitted energy is relatively small, $\Delta E \approx 0.00004$, and it requires many bounces to lower the energy below the annihilation threshold which is twice of the soliton mass. In the right panels the ${\rm K}\bar{\rm K}$ collision is plotted. Now, after the first bounce, the radiation takes a significant amount of energy, $\Delta E \approx 0.022$, which brings the solitons very close to the annihilation threshold. This confirms that the energy is radiated out in a very efficient way and, consequently, is an important factor in the dynamics. 

The first step in the construction of a correct CCM is the right choice of the one-parameter restricted set of configurations which interpolates between the initial and final states. In this case, the initial state, i.e., the infinitely separated pair of kink $\Phi_K$ and antikink $\Phi_{\bar{K}}$ located at $\mp a$ respectively, can produce a final state consisting of a pair of mirror antikink $\Phi_{\bar{K}}^*$ and mirror kink $\Phi_K^*$ located at $\mp a$. Schematically it can be represented as the following process $(0,1)+(1,0) \to (0,-1)+(-1,0)$.  Surprisingly, the usual naive superposition 
\be
\mathcal{M}^{naive}_{K\bar{K}} [a]= \{ \Phi_K(x;-a) + \Phi_{\bar{K}}(x;a) -1\} \label{kak-naive}
\ee
fails. Topologically, changing the modulus from $a=\infty$ to $a=-\infty$ gives a correct transition. In the initial state $(a\to \infty)$ we do have an infinitely separated kink-antikink pair, with a profile starting from 0 at $x=-\infty$, tending to 1 at $x=0$ and then decreasing to 0 for $x=\infty$. In the final state ($a \to - \infty$) we get a configuration which again tends to 0 for $x=\mp \infty$ and takes $-1$ at $x=0$. Although this is a correct topological behaviour, the final state is not a pair of mirror solitons,
\bea
\lim_{a\to -\infty} \left( \Phi_K(x;-a) + \Phi_{\bar{K}}(x;a) -1\right) &\neq &  \\
\Phi^*_{\bar{K}}(x;-\infty) &+& \Phi^*_K(x;\infty)+1 \nonumber
\eea
Hence, the assumed naive sum does not provide the correct final state. 

\begin{figure}
 \includegraphics[width=1.00\columnwidth]{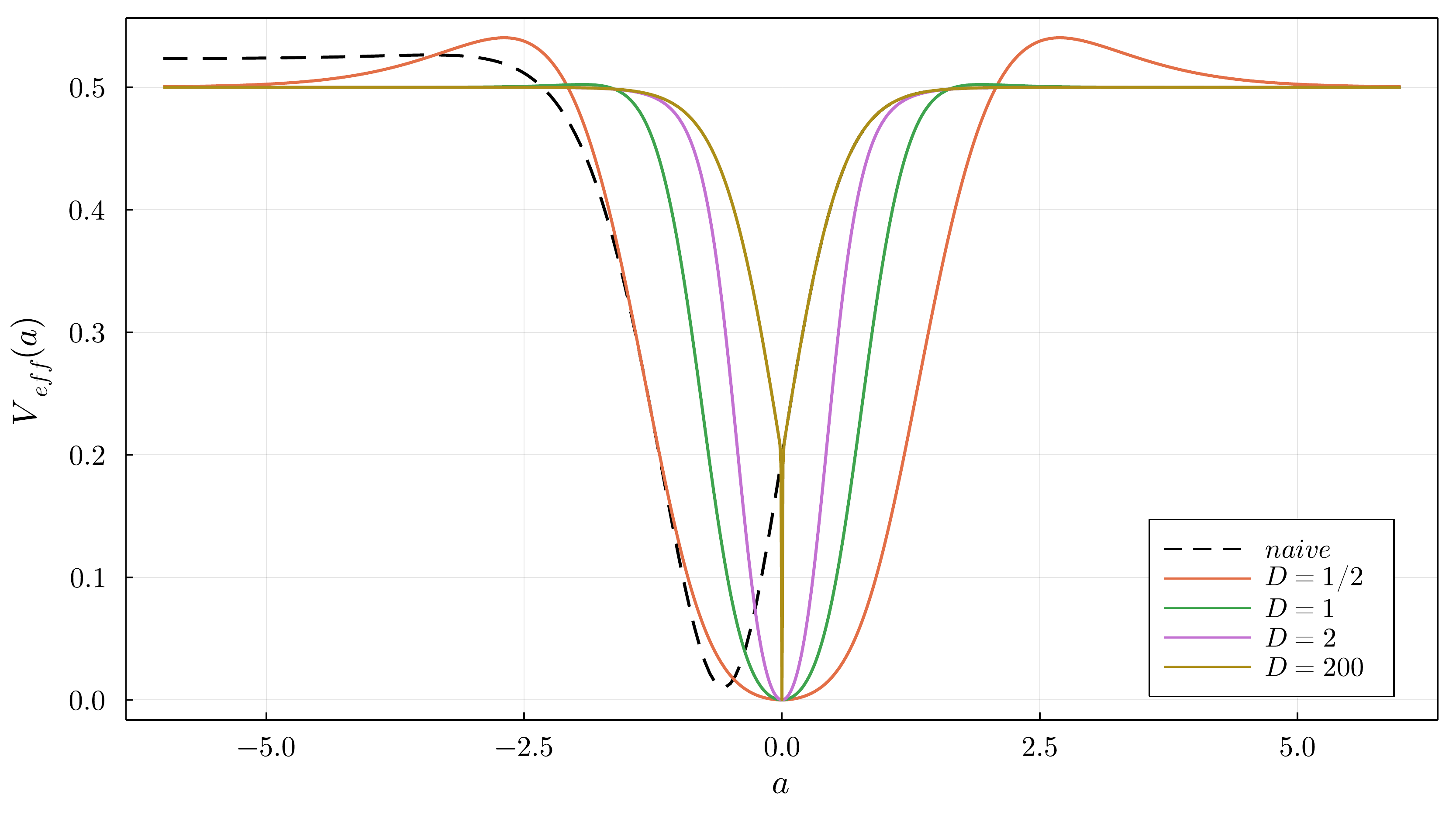}
  \caption{The effective potential in the one-dimensional CCM for ${\rm K}\bar{\rm K}$ collisions: dashed curve - the naive sum (\ref{kak-naive}); orange, green and purple curves - the improved choice with $D=0.5, 1, 2,200$.}\label{V_KAK}
     \end{figure}

Specifically, the large $|x|$ asymptotic of the naive sum (\ref{kak-naive}) for $a \to -\infty$ does not agree with the asymptotic of a largely separated mirror antikink and mirror kink. This leads to slightly different asymptotical values of the effective potentials in the corresponding CCM. Specifically, $V_- > V_+$, where $V_{\pm} = \lim_{a\to \pm \infty} V(a)$, see Fig. \ref{V_KAK} dashed curve. This unphysical jump in the CCM potential can have an unwanted impact on the resulting dynamics. 

A restricted one-parameter set of configurations which provides the correct initial and final states can have the following form
\bea
&&\Phi_{K\bar{K}}(x;a)= \label{KAK-improved}\\
&&\tanh(Da)  \left( \Phi_K(x;-a\tanh(Da)) + \Phi_{\bar{K}}(x;a\tanh(Da)) -1\right) \nonumber 
\eea
or explicitly
\bea
\Phi_{K\bar{K}}(x;a)&=& \tanh(Da) \left(   \sqrt{\frac{1+\tanh (x+a\tanh(Da))}{2}}+ \nonumber  \right.\\
&& \left. \sqrt{\frac{1-\tanh (x-a\tanh(Da))}{2}}-  1\right), 
\eea
where, for the moment, $D$ is a free constant. It is easy to verify that this set interpolates between the right initial and final states. Indeed,
\be
\lim_{a\to \infty} \Phi_{K\bar{K}}(x;a) = \lim_{a\to \infty}  \left( \Phi_K(x;-a) + \Phi_{\bar{K}}(x;a) -1\right) 
\ee
which is an infinitely separated pair of kink and antikink. Similarly, 
\bea
\hspace*{-0.4cm} \lim_{a\to -\infty} \Phi_{K\bar{K}}(x;a) &=& \lim_{a\to -\infty}  \left( -\Phi_{K}(x;a) - \Phi_{\bar{K}}(x;-a) +1\right) \nonumber \\
 &=&\lim_{a\to \infty}  \left( \Phi_{\bar{K}}^*(x;-a) + \Phi^*_{K}(x;a) +1\right)\hspace{-0.1cm},
\eea
which is an analogous pair of mirror antikink and mirror kink. This choice has the additional advantage that these states are approached exponentially fast. Thus, the naive superposition, both in the normal and mirror sector, is deformed only for small $a$. 

Note also that our configurations are antisymmetric in $a$, $\Phi_{K\bar{K}}(x;-a)=-\Phi_{K\bar{K}}(x;a)$. 

Since (\ref{KAK-improved}) provides the correct initial and final states, the corresponding effective potential has the same asymptotical values. Hence, there is no unwanted jump. However, the effective potential can still have a local maximum which again may lead to unwanted one-bounce windows. This in fact happens for $D \lesssim 1.27$, see Fig. \ref{V_KAK}. For bigger values the potential has only the global minimum for $a=0$, $V(a=0)=0$, where we pass through the vacuum $\Phi_v=0$.  

Of course, a CCM based on one collective DoF cannot explain the dynamics even in the crudest way. Here, however, there are no soliton-confined or delocalized normal modes whose amplitudes could provide new collective coordinates. Fortunately, the perturbative relativistic moduli space framework supplies us with the required additional collective excitations, which are Derrick modes. Thus, the simplest two dimensional moduli space with only one Derrick mode reads
\bea
&&\mathcal{M}_{K\bar{K}}[a,c]=\{ \Phi_{K\bar{K}}(x;a)  + c \tanh (Da) \times  \\
&& \left(\frac{a \tanh (Da)-x}{\left(e^{-2 (x-a \tanh (Da))}+1\right) \sqrt{e^{2
   (x-a \tanh (Da))}+1}}\right. \nonumber \\
   && \left. \left.+\frac{a \tanh (Da)+x}{\sqrt{e^{-2 (a \tanh (Da)+x)}+1} \left(e^{2
   (a \tanh (Da)+x)}+1\right)}\right)  \right\} \nonumber
\label{kak-pert}
\eea
where $\Phi_{K\bar{K}}(x;a)$ is the improved one-parameter configurations (\ref{KAK-improved}) and $c$ is the amplitude of the corresponding superposition of the modified first Derrick mode obtained by a scaling perturbation of (\ref{KAK-improved}). 

It is easy to notice that this restricted set of configurations suffers from a null vector problem. Indeed, $\partial_c\Phi(x;a=0,c)=0$. Therefore, $g_{ca}=g_{cc}=0$ at $a=0$. Fortunately, this singularity can be removed by a redefinition of the collective coordinate $c$ \cite{MORW}, \cite{AMORW}. Namely,
\be
c \to \frac{c}{\tanh (Da)}.
\ee
Thus, finally, the regular two-dimensional moduli space is defined as follows
\bea
&&\mathcal{M}_{K\bar{K}}[a,c]=\{ \Phi_{K\bar{K}}(x;a)  +  \label{2dim-improved} \\
&&c \left(\frac{a \tanh (Da)-x}{\left(e^{-2 (x-a \tanh (Da))}+1\right) \sqrt{e^{2
   (x-a \tanh (Da))}+1}}\right. \nonumber \\
   && \left. \left.+\frac{a \tanh (Da)+x}{\sqrt{e^{-2 (a \tanh (Da)+x)}+1} \left(e^{2
   (a \tanh (Da)+x)}+1\right)}\right)  \right\}. \nonumber
\label{kak-pert}
\eea

The optimal value of the parameter $D$ can be determined by comparison of the configurations (\ref{2dim-improved}) with the actual profiles observed in the scattering process. We found that $D=2$ is a good choice. We also remark that all $D \in (1,5)$ work similarly well and no big changes are observed. 

\begin{figure}
 \includegraphics[width=1.00\columnwidth]{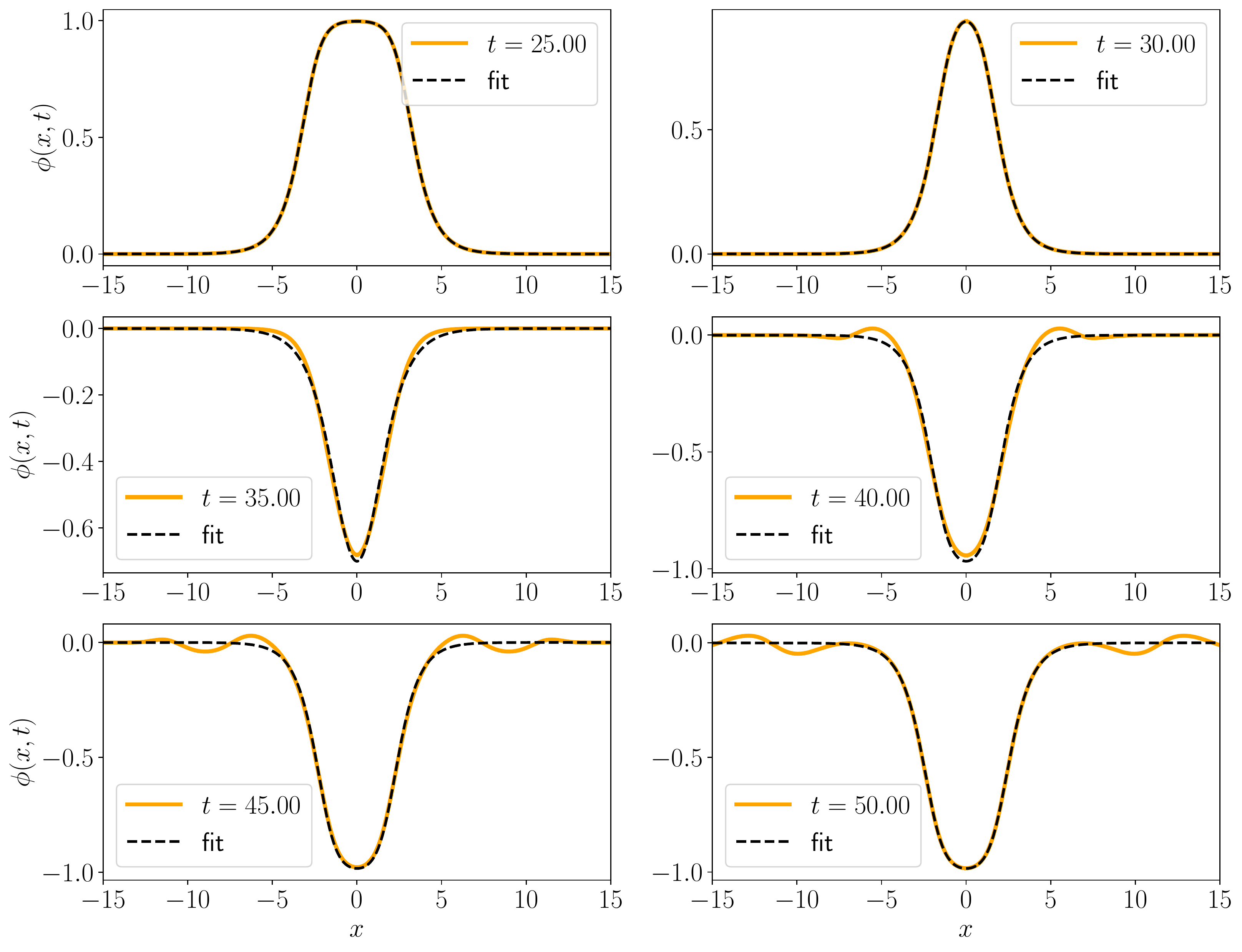}
  \caption{Comparison of true field profiles in a ${\rm K}\bar{\rm K}$ collision with $v_{in}=0.288$ with the best fit in terms of the two dimensional moduli space (\ref{2dim-improved}).}\label{AK_K_fit:fig}
     \end{figure}
\begin{figure}
 \includegraphics[width=1.00\columnwidth]{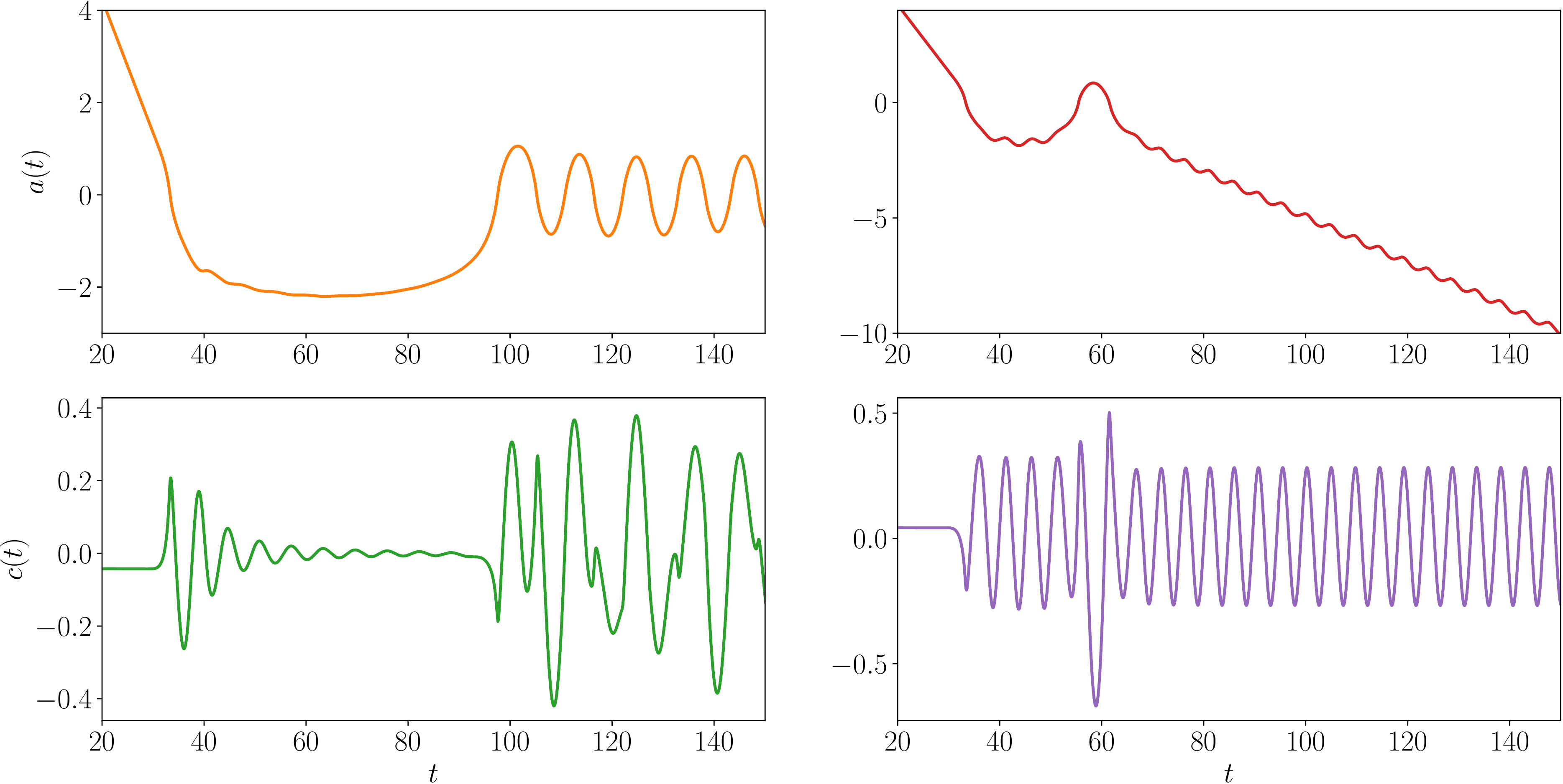}
  \caption{{\it Left:} Time dependence of the best fit of the moduli $a, c$ for the ${\rm K}\bar{\rm K}$ collision with $v_{in}=0.288$. {\it Right:} $a(t), c(t)$ computed from the CCM based on  (\ref{2dim-improved}). }\label{AK_K_fit_ac:fig}
     \end{figure}
     
In Fig. \ref{AK_K_fit:fig} we show the actual field profiles (solid line) together with the best fit for the two-dimensional moduli space configurations (dashed line) for several time moments. Here again $v_{in}=0.288$ and we demonstrate the situation close to the first collision. It is clearly visible that the configurations (\ref{2dim-improved}) very well approximate the field profiles except the inability to model the radiation far away from the center of the solitons. This is, of course, expected. Furthermore, once again we see how big is the radiation emitted during the collision. 

In Fig. \ref{AK_K_fit_ac:fig}, left panels, we show how the fitted values of the moduli $a$ and $c$ change in time. This is not a CCM result but a fit of the configurations  (\ref{2dim-improved}) to the true profiles obtained in the full partial differential evolution. We see a relatively long first window and fast decay of the excited first Derrick mode after the first collision. This agrees with the emission of radiation which quickly releases energy from the soliton-confined DoF. The late time chaotic changes of the amplitude of the Derrick mode probably imply that the moduli (\ref{2dim-improved}) does not work too well and some other DoF must be included. 

Now we show the results obtained in the CCM based on the moduli (\ref{2dim-improved}). The initial conditions are again 
\be
a(0)=a_0, \;\; \dot{a}=v_{in}, \;\; c(0)=\tilde{c}, \;\; \dot{c}=0
\ee
where $\tilde{c}$ is a stationary solution of the single soliton case, exactly as for the $\bar{\rm K}{\rm K}$ collision in pRCCM. In our analysis this is computed numerical for each $v_{in}$. 

\begin{figure}
\hspace*{-0.5cm}  \includegraphics[width=1.15\columnwidth]{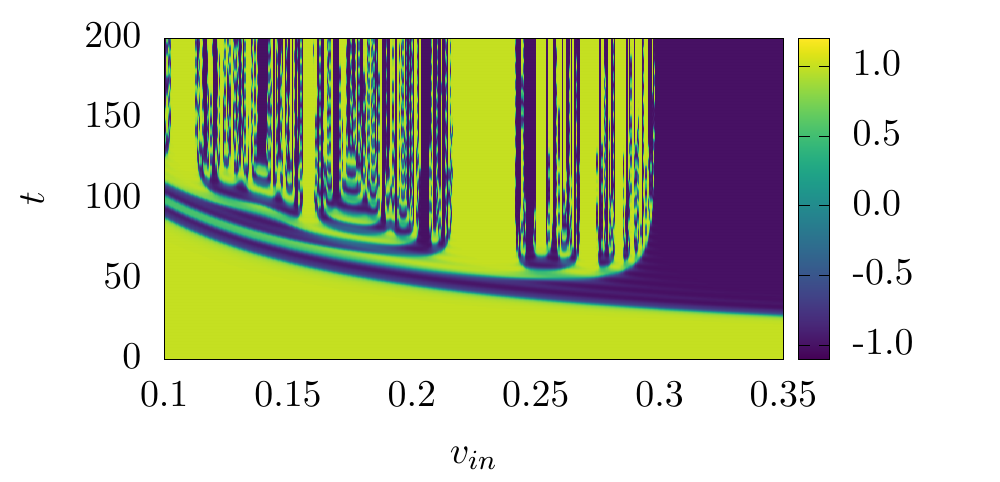}
  \caption{${\rm K}\bar{\rm K}$ collisions in the $\phi^6$ model in a CCM based on the moduli (\ref{2dim-improved}). We plot the time dependence of the field at the origin as a function of the initial velocity $v_{in}$.}\label{KAK-CCM:fig}
     \end{figure}

In Fig. \ref{KAK-CCM:fig} we show the time dependence of the value of the field at the origin for initial velocities $v_{in} \in [0.1,0.35]$. An encouraging result is that this simple CCM reproduces the critical velocity quite well. Namely, we found $v_{cr}^{CCM}\approx 0.298$ which is quite close to the true value $v_{cr}\approx 0.289$. On the other hand, the whole structure of the formation of the final state is not accurately predicted by the CCM. For example, besides the nice bion chimneys, we also see quite well pronounced, unwanted multi-bounce windows. The origin of the appearance of these unwanted features is quite clear. The CCM has only soliton confined DoF and, therefore, there is no way to dissipate energy located on a kink or antikink. As a consequence, during the CCM evolution the energy stored in the first Derrick mode can at later times be back-transferred to the kinetic motion, which eventually leads to back-scattering of the solitons. This, of course, does not happen in the full field theoretical collision, where the radiation quickly reduces the energy of the kinks, leading to a complete annihilation. 

The correct prediction of the critical velocity may suggest that the first Derrick mode is good enough to  describe the early phase of the collision properly, where it is decided which amount of the energy is transferred from kinetic motion to other DoF and, therefore, it is decided whether we have a simple passing-through collision, or we enter in a more complicated scenario. This is in fact visible if we plot $a(t)$ and $c(t)$ obtained from the CCM and compare it with the best fit, see Fig. \ref{AK_K_fit_ac:fig}, right panels. We found a reasonable good agreement for the first collision. Namely, we see a wider first bounce window. Also the amplitude of the excited Derrick mode agrees. However, as there is no dissipation, this amplitude remains basically constant. This has a significant impact on the late time behaviour and leads to the observed disagreement with the full theory. 

%%%%%%%%%%%%%%%%%%%%%%%%%%%%%
\section{Conclusions}
%%%%%%%%%%%%%%%%%%%%%%%%%%%%%

In this work, we fully confirmed that the fractal structure observed in the final state formation in antikink-kink collision in the $\phi^6$ model is caused by delocalized, two-soliton modes. These modes arise in the linear perturbation theory of the antikink-kink static configuration and, contrary to the usual shape modes, are not hosted by a single soliton solution. So, they are non-perturbatively related to the antikink-kink sector. As a consequence, the properties of these delocalized modes (their shapes and frequencies) as well as their number depends on the inter-soliton distance. 

This was confirmed by a close analysis of the actual field profiles as well as by a construction of a simple two-dimensional collective coordinate model (CCM) which {\it amazingly well} reproduces the full theory computations. We would like to underline that the results obtained in the CCM agree with the full field theory with a precision which goes significantly beyond previously presented collective approximations such as, for example, kink-antikink scattering in the $\phi^4$ model \cite{MORW}. 

Based on the success of the CCM description of the $\bar{\rm K}{\rm K}$ collisions in the $\phi^6$ model (and the ${\rm K}\bar{\rm K}$ collision in $\phi^4$ theory) we propose a general, robust way for the construction of CCM for a wide class of multikink processes. The main ingredient is the restricted set of configurations (parametrized by a finite set of collective coordinates) which includes: {\it (1)} a one-parameter subset configurations interpolating between the initial and final states of the process. This is typically, but not always, given by the naive superposition of static solutions of solitons which participate in the process; {\it (2)} the linear modes arising in the {\it multi-soliton} linear perturbation theory; and finally {\it (3)} Derrick modes which take into account the Lorentz contraction of the constituent solitons and also may serve to model some effects of radiation. These components should provide a CCM which may be treated as a precision tool allowing for a {\it quantitative} understanding of many kink collisions. This approach should lead to the explanation of fractal structures found in multikink collisions in various models as, for example, $\phi^8$ theory or other models having solitons with so-called fat tails. We expect that this should work especially well whenever there is a (single- or multi-soliton) normal mode involved in the process. 

If no such modes are present, the situation is surprisingly more subtle, because the main factor governing the dynamics is the interaction of kinks with radiation. Unfortunately, the inclusion of radiative (not soliton confined) DoF into a collective model framework is still not fully developed. However, the Derrick modes may provide a simple and useful approximation. This follows from the observation that for sufficiently high Derrick modes the frequencies begin to be higher than the mass threshold. Furthermore, higher modes are more widespread and therefore give a better approximation to more distant regions. Of course, the main limitation is that these modes are also soliton confined and cannot transfer energy to infinity. 

Thus, although the proposed general scheme for the construction of CCMs is based on discrete DoF ((quasi)normal modes, Derrick modes), it does take into account some feature of radiation, which is also an important step forward. Interestingly, delocalized modes can also be viewed as {\it localized radiation}, i.e., a kind of standing wave trapped between the colliding solitons. This includes only radiation with frequency lower than the mass threshold of the outer vacuum. In addition, Derrick modes also enjoy some features of radiation, especially at a short time scale. This we saw in a quite accurate value of the critical velocity for the ${\rm K}\bar{\rm K}$ collision. 

\vspace*{0.2cm}

 There are many directions in which the current work should be continued besides the straightforward application to multikink processes in other single scalar field theories.  

Firstly, we need a CCM where the delocalized, two-soliton modes are treated dynamically, i.e., without any freezing procedure. As we already mentioned, this a rather nontrivial task which, in its full shape, might require to incorporate threshold modes, anti-normal modes and quasi-normal modes, i.e., the modes into which the delocalized normal modes transmute as the solitons approach each other. This would allow us to get rid of the free parameter $a_*$ which enters the current version of the CCM for the $\bar{\rm K}{\rm K}$ collision. 

Secondly, especially in the cases where there is no normal mode involved in the process (as in the ${\rm K}\bar{\rm K}$ collision in $\phi^6$ theory) one should couple a bigger number of Derrick modes. This should allow for a better modeling of the effects of radiation, which in this case is the crucial factor. We expect that the addition of higher Derrick modes will improve the medium time CCM dynamics by introducing a channel in with the energy can be transferred in quite an efficient way. This should lead to the disappearance of unwanted bounce windows which exist in our CCM, leading to a much better accordance of the CCM with the full theory. 

Another issue is related to the construction of the improved one-dimensional moduli space which replaces the invalid usual naive sum. It would be desirable to find a unique and general way for such a construction. 

%%%%%%%%%%%%%%%%%%%%%%%%%%%%%%%%%%%%
\section*{Acknowledgements}
%%%%%%%%%%%%%%%%%%%%%%%%%%%%%%%%%%%%

The 
authors acknowledge financial support from the Ministry of Education, Culture, and Sports, Spain (Grant No. PID2020-119632GB-I00), the Xunta de Galicia (Grant No. INCITE09.296.035PR and Centro singular de investigación de Galicia accreditation 2019-2022), the Spanish Consolider-Ingenio 2010 Programme CPAN (CSD2007-00042), and the European Union ERDF.
AGMC is grateful to the Spanish Ministry of Science, Innovation and Universities, and the European Social Fund for the funding of his predoctoral research activity (\emph{Ayuda para contratos predoctorales para la formaci\'on de doctores} 2019). MHG thanks the Xunta de Galicia (Consellería de Cultura, Educación y Universidad) for the funding of his predoctoral activity through \emph{Programa de ayudas a la etapa predoctoral} 2021. KO was supported by the Polish National Science Centre 
(Grant NCN 2021/43/D/ST2/01122). The research of PED was supported in part by the National Science Foundation under Grant No.\ NSF PHY-1748958, and in part from the STFC under consolidated grant ST/T000708/1.

%\newpage

\end{document}